\documentclass[a4paper,fleqn]{cas-sc}

 \usepackage[numbers]{natbib}
\usepackage{enumitem}  
\usepackage{float}
\usepackage{graphicx}
\usepackage{tabularx}
\usepackage{ltablex}     
\usepackage{multirow}
\usepackage{booktabs}    
\usepackage{fontawesome5} 
\usepackage{pdflscape}
\usepackage{listings}
\usepackage{xcolor}
\usepackage{graphicx}
\usepackage{subcaption}
\usepackage{calc}  
\usepackage[most]{tcolorbox}

\definecolor{clrsupported}{HTML}{C6EFCE}
\definecolor{clrpartial}{HTML}{FFEB9C}
\definecolor{clrrevision}{HTML}{FFCCCC}
\definecolor{clrnoeval}{HTML}{DDDDDD}
\definecolor{clrsummary}{HTML}{DCE8F0}
\definecolor{clrgrphead}{HTML}{EEF4FB}

\tcbset{
  colback=white,
  colframe=black,
  boxrule=0.5pt,
  arc=2pt,
  left=6pt,
  right=6pt,
  top=4pt,
  bottom=4pt,
  breakable, 
  enhanced
}

\setlist[enumerate]{noitemsep, topsep=0pt}
\setlist[itemize]{noitemsep, topsep=0pt}

\def\tsc#1{\csdef{#1}{\textsc{\lowercase{#1}}\xspace}}
\tsc{WGM}
\tsc{QE}
\tsc{EP}
\tsc{PMS}
\tsc{BEC}
\tsc{DE}

\begin{document}
\let\WriteBookmarks\relax
\def\floatpagepagefraction{1}
\def\textpagefraction{.001}

\shorttitle{Human-Centered RE for Critical Systems}


\title [mode = title]{Human-Centred Requirements Engineering for Critical Systems: Insights from Disaster Early Warning Applications}                      



\author[1]{Anuradha Madugalla}

\cormark[1]

\fnmark[1]

\ead{anuradha.madugalla@deakin.edu.au}



\affiliation[1]{organization={ School of Information Technology, Deakin University},
   city={Melbourne},
     state={Victoria},
   country={Australia}}

\author[2]{Jixuan Dong}
\author[2]{Kai Lyne Loi}
\author[2]{Matthew Crossman}
\author[2]{John Grundy}

\affiliation[2]{organization={Faculty of Information Technology, Monash University},
   city={Melbourne},
     state={Victoria},
   country={Australia}}

\begin{abstract}
Critical systems, such as those used in healthcare, defence, transportation, and disaster management, require rigorous requirements engineering to ensure safety and reliability. However, this rigour has traditionally focused on technical assurance, with less attention to the human and social contexts in which these systems are used. This paper argues that human-centricity is an essential dimension of dependability and presents a human-centred requirements engineering process for making vulnerable-user needs explicit and traceable from inclusive design guidelines to requirements, prototype features, and validation evidence. Drawing on a structured review of inclusive design literature, we identified 62 guidelines relevant to four vulnerable communities: older adults, low-digital-literacy users, rural users, and colour-blind users. These guidelines were translated into a catalogue of 67 functional and non-functional requirements for inclusive early warning systems. To our knowledge, this is one of the first studies in software engineering to consolidate and empirically validate inclusive design requirements for disaster early warning systems, where accessibility and usability failures can have serious safety consequences. The requirements were operationalised through an adaptive disaster early warning prototype and evaluated through six interviews and eight cognitive walkthroughs. The evaluation provided strong positive evidence across the four vulnerable groups, with particularly encouraging results for elderly and rural users, whose requirements achieved full validation coverage and high positive validation rates. The paper concludes by positioning human-centricity not as an ethical add-on, but as a traceable quality concern in the design of safe and equitable critical systems.
\end{abstract}


\maketitle

\section{Introduction}
Critical systems, such as those used in healthcare, transportation, defence, and disaster management, demand exceptional rigour in their requirements engineering (RE). When these systems fail, the consequences extend beyond technical malfunction; they directly threaten human safety and public trust \cite{martins2016}. Therefore, RE for critical systems has historically focused on aspects such as safety, dependability, and correctness. While these qualities remain essential, this emphasis has often come at the expense of an equally critical dimension: the integration of human-centred concerns. 

Neglect of these concerns can lead to significant consequences. A healthcare system that assumes expert digital literacy of its users may exclude elderly patients; an emergency transport app optimised for urban connectivity may fail rural users; and a disaster early warning system designed for the “average” user may leave people with disabilities without access to life-saving alerts. In critical systems, these shortcomings cannot be considered as simple usability oversights. Instead, they must be recognised as requirements failures rooted in incomplete stakeholder analysis and non-inclusive elicitation practices.

Focusing on inclusive elicitation not only supports vulnerable communities but also leads to broader improvements in product quality and user experience for everyone. For example, in the broader field of engineering, curb cuts were created to make city streets more accessible for wheelchair users. But this resulted in curb cuts enhancing mobility for many others, including people pushing strollers, pulling suitcases, delivering goods, or riding bicycles and skateboards \cite{churchill2018}. A similar pattern can be observed in software inclusivity: captions were first introduced to make videos accessible to people with hearing impairments, but with the rise of social media, they are now used widely \cite{hong2010}. Captions enable quiet or multilingual viewing, showing how accessibility features can evolve into mainstream usability improvements \cite{gernsbacher2015video}. These examples illustrate that inclusive design is not a constraint but a driver of innovation and universal benefit. However, a significant gap exists in ensuring that requirements capture the full spectrum of stakeholder needs, particularly those of vulnerable or marginalised communities \cite{sutcliffe2004}. Research has consistently shown that RE in safety-critical domains remains largely compliance-focused, with limited mechanisms for engaging diverse stakeholders or addressing the needs of vulnerable user populations \cite{martins2016, martins2020, hidellaarachchi2021}.

This paper builds on the argument that socially responsible RE is central to the development of all systems, especially for critical systems. We propose an RE process that ensures requirements are elicited, specified, and validated in ways that support social equity. To demonstrate this process, we present a case study in disaster management: the design of an adaptive early warning application that addresses the needs of vulnerable groups. We show how participatory engagement with vulnerable communities can uncover overlooked requirements, translate them into adaptive design guidelines, and validate them through stakeholder feedback. Our contributions are threefold:

\textit{\textbf{Contribution 1- A validated catalogue of inclusive requirements: }}
We contribute a catalogue of 67 functional and non-functional requirements for designing inclusive early warning systems for four vulnerable communities: older adults, low-digital-literacy users, rural users, and colour-blind users. These requirements were derived from 62 inclusive design guidelines and evaluated through stakeholder-validated prototypes. To our knowledge, this is one of the first consolidated requirements catalogues developed specifically for inclusive early warning systems, where failures in accessibility, comprehension, or actionability can have serious safety consequences.

\textit{\textbf{Contribution 2- A traceable human-centred RE process for inclusive safety-critical systems: }}
We contribute a replicable human-centred RE process that connects evidence-based elicitation, structured requirements specification, prototype-supported operationalisation, and participatory validation. Rather than treating inclusivity as a late-stage usability concern, the process makes vulnerable-user needs explicit and traceable from design guidelines to requirements, prototype features, and validation evidence. We demonstrate the process through an end-to-end case study in disaster early warning and discuss how it can inform other user-facing safety-critical domains.

\textit{\textbf{Contribution 3: Empirical insights into the validation of human-centred requirements:} }
We provide empirical evidence from the prototype evaluation showing how human-centred requirements can be confirmed, refined, or left unevaluated, depending on the prototype's capabilities and the validation method. Across the 67 requirements, 84\% were addressed during validation, and 70\% received positive validation evidence. The findings show that addressing human-centric requirements early can improve  usability for vulnerable communities, while also revealing important trade-offs around adaptivity, colour customisation, connectivity, and implementation-dependent requirements.

\section{Motivation}

\begin{figure}[h]
\centering\includegraphics[width=0.75\textwidth, trim={0.3cm 38cm 2cm 0cm},clip]{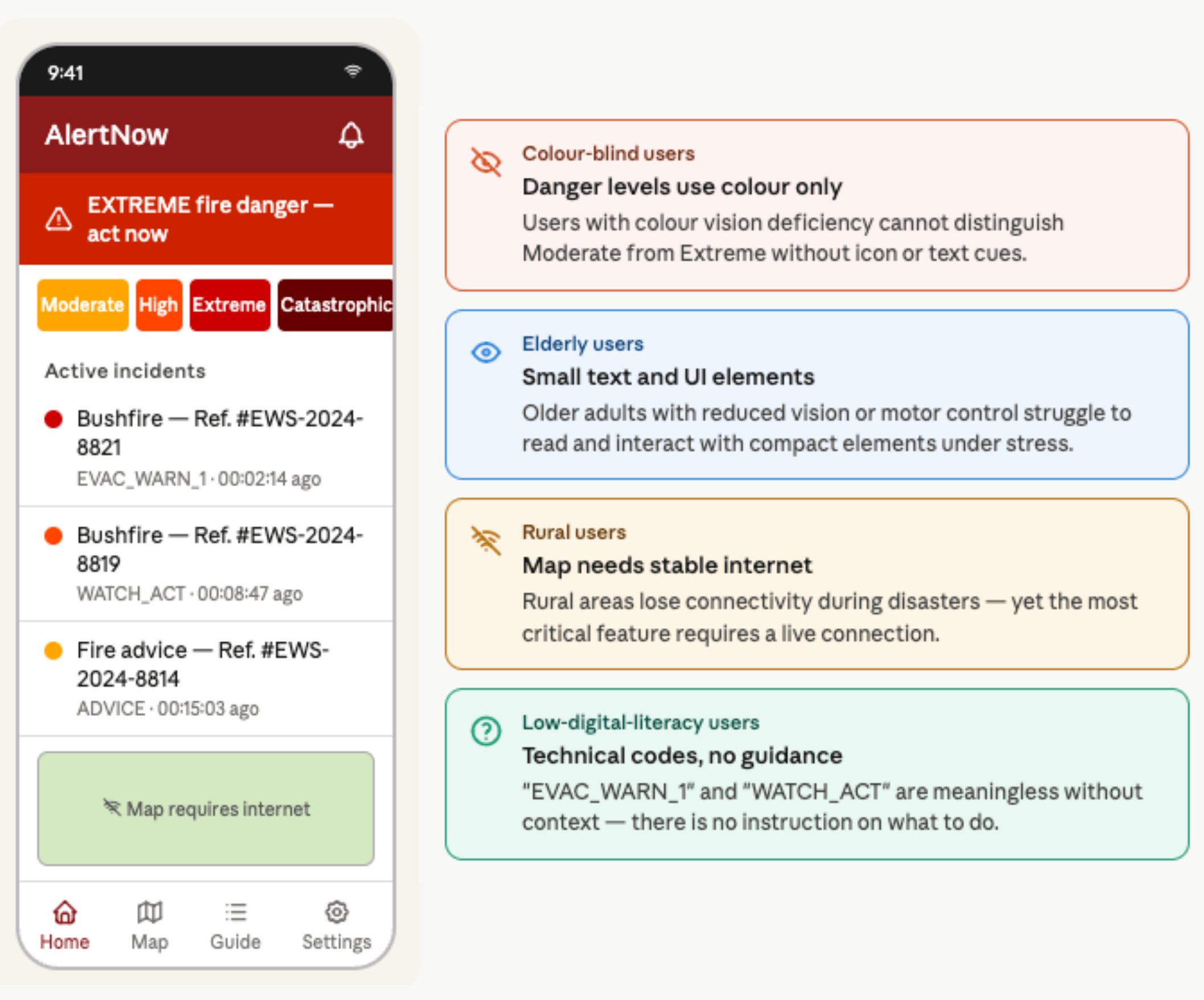}
\caption{An System Designed for an `Average' User Failing Those Who Need it Most}
\label{fig_motivation}
\end{figure}

Software is designed to address human needs. Yet, we continue to see systems that fail to meet the needs of some intended users, particularly when they are designed around a notional “average” user as shown in Figure \ref{fig_motivation}. This one-size-fits-all approach overlooks the needs of vulnerable groups. For example, in a disaster warning system, colour-only alerts may be inaccessible to users with colour vision deficiencies; small interface elements may exclude older adults with vision issues; and reliance on stable connectivity disadvantages rural populations. Such oversights hinder their access to critical early warnings, placing their lives and livelihoods at danger. When warnings cannot be understood or acted upon by the most vulnerable, the system fails in its primary purpose. Such oversights demonstrate the necessity of embedding human-centricity into requirements engineering for critical systems.

This study is motivated by the need to bring human-centricity and participatory principles into the early phases of RE for critical systems. By deriving requirements from participatory design guidelines, modelling them through adaptive prototypes, and validating them with both communities and personas, this research demonstrates a pathway towards human-centred RE that is both socially responsible and technically rigorous. Ultimately, it aims to ensure that no user is left behind due to a lack of representation in the system development process. To achieve this, the paper investigates the following research questions:

\textit{RQ1: What are the requirements of vulnerable groups in the context of critical systems?}

\textit{RQ2: How can these requirements be modelled and operationalised for stakeholder feedback?}

\textit{RQ3: How can requirements be validated with diverse stakeholders to ensure their needs are effectively addressed?}

To investigate these research questions, we focus on disaster early warning systems as the case domain. These are safety-critical systems where access failures can have severe consequences, and where vulnerable communities are often disproportionately affected.

The rest of the paper is structured as follows. In the next section, we present a review of related work. Section 4 discusses our human-centred requirements engineering process, and sections 5 and 6 contain two steps of this approach: requirements elicitation and specification. Section 7 presents how these requirements were operationalised through prototyping, and Section 8 presents the prototype evaluation and stakeholder feedback. Section 9 discusses the findings, and Section 10 presents threats to validity. The final section of the paper summarises the study, presenting suggestions for future work.

\section{Related Work}

\subsection{Requirements Engineering for Critical Systems}
Requirements Engineering in critical systems has long focused on aspects such as safety, reliability and correctness. Within domains such as healthcare, transportation, defence, and disaster management, these qualities are vital to prevent major failure and to ensure regulatory compliance. This was brought forward by early foundational work, such as that by Van Lamsweerde, which emphasised that correctness and traceability are qualities that are central to the development of critical systems \cite{vanlamsweerde2000}. Over time, these principles have strongly influenced safety system-oriented RE frameworks and compliance standards across critical domains.

However, despite decades of methodological development, safety-critical RE remains strongly technically oriented, which has left human and social dimensions of RE largely unaddressed. Hatcliff et al. identify establishing complete, correct, unambiguous, and understandable requirements as one of the biggest unsolved challenges in safety-critical systems engineering, and observe that system engineering, safety engineering, and software engineering remain poorly integrated disciplines \cite{hatcliff2014certifiably}. Similar concerns have been raised by Heimdahl \cite{heimdahl2007safety} and Leveson \cite{leveson1995safeware} over several decades, suggesting that this fragmentation remains a persistent issue in the field.

This technical orientation is also reflected in the research base. Martins and Gorschek's systematic review of 151 studies found that only 7.28\% of the studies proposed approaches that supported both safety analysis and requirements engineering in a genuinely cooperative, integrated way \cite{martins2016}. The review also showed that traditional approaches, such as FTA, FMEA and HAZOP, continue to dominate, having spread from their origins in the military and chemical industries in the 1960s to become effectively universal across safety-critical domains. In contrast, newer approaches have struggled to gain traction: 73 of 151 reviewed studies proposed approaches that appeared in only a single study and were never developed further; most were proposed without industry collaboration and without addressing topics practitioners actually consider relevant.

This technical orientation is reinforced by regulatory pressure. Certification agencies require accurate, traceable requirements specifications, and requirements deficiencies are widely understood in the safety-critical systems community as the biggest source of unanticipated cost and delivery delays \cite{kornecki2009certification}. As a result, requirements documents are often produced to satisfy compliance expectations \cite{martins2016}. While this is necessary, it can narrow the role of RE to technical assurance and documentation, rather than using it as a process for understanding diverse stakeholder needs throughout the system lifecycle.

The consequences of this orientation are visible in both research and practice. Martins and Gorschek found that 95\% of the reviewed studies provided no meaningful evidence of how the usability of proposed approaches was measured \cite{martins2016}. This is not simply a methodological gap; it reflects a deeper assumption embedded in safety-critical RE research that technical correctness and regulatory compliance are sufficient criteria for evaluating an approach, while the experiences of the people who must use, understand, and act on these systems remain largely invisible. Martins and Gorschek's interview study with 19 practitioners from 11 safety-critical companies confirmed this pattern in practice \cite{martins2020}. Requirements were routinely elicited through informal interviews and unstructured meetings rather than systematic methods. Sixty-one percent of companies used only word processors and spreadsheets to manage their requirements. No company had requirements engineers participating regularly in safety teams. Practitioners described requirements documents as primarily produced to demonstrate compliance rather than as genuine communication tools throughout the development lifecycle. User engagement and feedback loops were minimal, and the needs of end-users, let alone vulnerable or marginalised users, were effectively absent from the requirements process.

In summary, existing work has made significant progress in providing technical assurance for safety-critical systems. However, it provides less guidance on how to address the human and social dimensions of these systems through RE. In particular, it remains unclear how the needs of diverse and vulnerable end users can be traced and systematically validated throughout the RE lifecycle, rather than deferred to late-stage usability testing or overlooked entirely. This motivates the need to examine human-centred and participatory RE approaches, which we discuss next.


\subsection{Human-Centred and Participatory Requirements Engineering}
Early work by Sutcliffe et al. demonstrated that neglecting human workload and task distribution can undermine both safety and performance, introducing the notion that human factors should be integrated into the RE process \cite{sutcliffe2004}. This idea later evolved into a broader recognition that RE is a socio-technical process that is shaped by values and relationships among diverse stakeholders. The systematic review by Hidellaarachchi et al. strengthened this perspective, showing that human aspects such as empathy, collaboration, and communication are central to RE success, and yet these remain under-represented in mainstream practice \cite{hidellaarachchi2021}. 

More recent research has extended this school of thought to encompass human aspects such as ethics and equity. Perera et al. empirically demonstrated that explicitly embedding human values—such as fairness, dignity, and social justice within the RE processes results in more creative and socially responsible requirements \cite{perera2021}. Their findings show that values can be operationalised through lightweight elicitation techniques. Similarly, Damian et al. redefined participatory RE as an ongoing, relationship-centred practice through the \textit{REConnect} framework \cite{damian2025}. Their work brings forward the idea that empathy, trust, and continuous collaboration are essential RE artifacts showing how long-term partnerships with communities produce requirements that are both technically sound and socially legitimate. Building on this trajectory, Tizard et al. called for more inclusive RE practices that consciously address the representation of underserved users \cite{tizard2024}. This work highlighted how both traditional elicitation methods and modern crowd-based approaches systematically underrepresent vulnerable groups, including people with disabilities, women, and older adults. The authors proposed practical strategies, such as combining diverse elicitation channels and employing tools like GenderMag cognitive-style surveys, to identify whose voices are missing in the process. However, they did not propose an end-to-end RE process to ensure human-centricity. 


Despite this progress, human-centred and participatory RE approaches have received limited systematic attention in safety-critical domains. This is especially true for intersectional vulnerability factors such as disability, ageing, digital literacy, and connectivity, which can directly affect users' ability to engage with these systems in high-stakes situations. Some work has addressed specific user groups, for example, Lockerbie and Maiden modelled quality-of-life goals for people living with dementia within an information systems context \cite{lockerbie2022modelling}, Wang et al captured requirements for people with cognitive impairments via their carers \cite{wang2026proxy}, but existing frameworks rarely incorporate multiple intersecting vulnerability factors in a way that is traceable from elicitation through to validated requirements. As Martins and Gorschek observed \cite{martins2016}, current industrial RE practices still marginalise diverse perspectives under procedural and regulatory constraints, and their SLR found that only 3 of 151 studies proposed approaches that explicitly addressed improving communication among all actors across the safety-critical systems' lifecycle. The needs of vulnerable end users, in particular, have received almost no attention in safety-critical RE research or practice. This study addresses this gap by deriving requirements from evidence-based inclusive design guidelines, operationalising them through an adaptive prototype, and validating them with community members and personas. It demonstrates how human-centricity can be embedded into the RE lifecycle for safety-critical systems in a traceable and replicable way.

\subsection{Inclusive Design and Accessibility}

Inclusive design and accessibility have been widely studied in the fields of Human-Computer Interaction (HCI) and software engineering. Existing work provides guidance for designing systems that better support users with diverse abilities and experiences \cite{shneiderman2021designing,gilbert2019inclusive,WebConte89_online}. For older adults, prior studies highlight the importance of readability, reduced cognitive load, simple navigation, and interaction patterns that account for age-related changes in vision, memory, and motor control \cite{gomez2023design,ruzic2017universal}. For low-literate and low-digital-literacy users, existing work emphasises plain language, familiar terminology, consistent layouts, step-by-step guidance, and error-tolerant interaction \cite{srivastava2021actionable,Designin57_online}. Research on rural digital inclusion has highlighted the need to consider connectivity constraints, low-bandwidth design, offline access, discoverability, and alternative information infrastructures \cite{groves2000web,roberts2017review,hardy2019designing}. For colour-blind users, existing guidance stresses the importance of avoiding colour-only communication, maintaining adequate contrast, using redundant visual cues, and testing designs with accessibility tools and colour-blindness simulators \cite{colorblindguideEssentialGuidelines,ColorBli89_online,WebConte89_online}. These studies provide a strong foundation for designing more inclusive software systems.

However, inclusive design guidance remains spread across several bodies of work. These include accessibility standards such as WCAG\cite{WebConte89_online}, platform and design-system guidelines such as the Apple Human Interface Guidelines\cite{HumanInt91_online}, Google Material Design accessibility guidance \cite{Accessib59_online}, and Microsoft’s Inclusive Design Toolkit \cite{Microsof23_online}, as well as practitioner-oriented usability guidance such as Nielsen Norman Group resources \cite{Accessib42_online}. In addition, HCI studies provide population-specific recommendations for groups such as older adults, low-digital-literacy users, rural users, and colour-blind users, as discussed above. While these sources provide valuable guidance for inclusive design, they are often not expressed as requirements artefacts that can be systematically traced and validated. This makes it difficult to apply such guidance during RE, especially when a system must support multiple vulnerable communities at once.

This distinction matters because design guidelines and requirements serve different purposes. Guidelines often describe broad design considerations, while requirements need to define what the system should do or how it should behave in a way that can guide development and evaluation \cite{chung2012non}. For example, a guideline such as ``avoid colour-only communication'' \cite{WebConte89_online,ColorBli89_online} becomes more actionable when translated into a requirement specifying that critical warnings should include redundant cues such as text, icons, patterns, or labels. Similarly, guidance on reducing cognitive load becomes more useful when expressed as requirements for simplified navigation, consistent interaction paths, and step-by-step instructions \cite{gomez2023design,srivastava2021actionable}. Translating guidelines into requirements, therefore, makes inclusive design concerns visible throughout the RE lifecycle and links them to source evidence and validation outcomes.

This is particularly important in user-facing critical systems where accessibility and usability failures are not simply matters of inconvenience. If users cannot access information, understand warnings, or act on instructions in time, the system may fail in its protective purpose \cite{martins2016}. Disaster early warning systems make this issue especially clear because they need to communicate urgent and often life-saving information to diverse users, including those who may experience sensory, cognitive, digital, geographic, or connectivity-related barriers \cite{tan_modified_2020,syukron2024comprehensive,madugalla2026role}. Therefore, inclusive design concerns should be treated as requirements-level issues, rather than as late-stage usability improvements.

Another limitation of existing guidance is that it often focuses on individual user groups. However, vulnerable users may experience overlapping needs. For example, simplified navigation may support older adults, low-digital-literacy users, and rural users, while high-contrast visual design may support both colour-blind users and older adults \cite{gomez2023design,srivastava2021actionable,WebConte89_online}. At the same time, some requirements may introduce trade-offs. For instance, colour customisation may improve accessibility for colour-blind users, but it may also increase complexity if users do not understand the available settings \cite{madugalla2025engineering}. These overlaps and tensions need to be identified during RE, rather than discovered only after deployment.

Building on this body of work, our study consolidates inclusive design guidance for four vulnerable communities and translates it into a traceable catalogue of functional and non-functional requirements for disaster early warning systems. In doing so, it connects HCI and accessibility knowledge with RE practice. The contribution lies in translating this knowledge into requirements, operationalising it, and validating it in a safety-critical context, so that vulnerable-user needs can be made explicit and traceable throughout the RE lifecycle.

\section{Methodology}
\begin{figure}[h]
\centering\includegraphics[width=\textwidth, trim={0.3cm 23cm 0.2cm 2cm},clip]{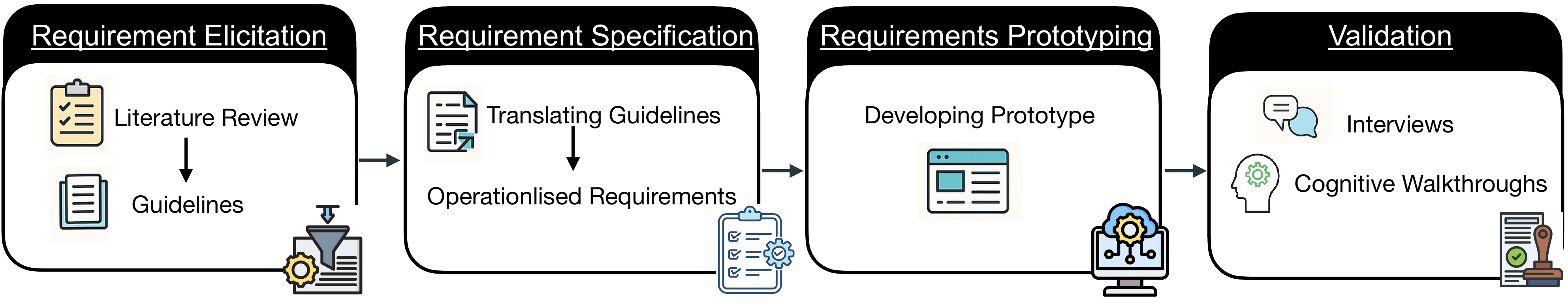}
\caption{Human-Centric Requirements Engineering Process}
\label{fig_method}
\end{figure}

In this study, we followed a four-stage requirements engineering process to ensure human-centric aspects of the requirements are captured (Figure \ref{fig_method}). 

The first stage was requirements elicitation, which focused on identifying the needs of our chosen vulnerable user groups. We conducted a targeted review of existing literature, including academic publications and grey literature, to extract guidelines for designing inclusive software relevant to the chosen groups: elderly, individuals with low digital literacy, rural residents, and people with colour-vision impairment.

The second stage was requirements specification, in which we interpreted and translated the extracted guidelines into implementable software requirements for safety-critical systems. Wherever possible, these requirements were expressed as functional or non-functional requirements with measurable criteria (for example, minimum button size, colour-contrast thresholds, or offline caching rules).

The third stage involved requirements prototyping, where the specified requirements were operationalised via a medium-fidelity interactive prototype. This prototype served as both a visualisation of the requirements and as a communication tool for engaging with stakeholders. For the prototype, we chose to design an Early Warning System (EWS) mobile application, which is commonly used in disaster communications to convey safety-critical information.   

Finally, the requirements validation stage assessed whether the prototype met stakeholder needs. This evaluation contained two steps: 1) Structured interviews with community members (six participants: two older adults and four rural residents), and 2) Cognitive walkthroughs using personas representing the four vulnerable groups. The insights we gained from both activities were used to validate the requirements and refine the prototype, ensuring that it addressed the needs of these communities.

A key highlight of our methodology is the integration of evidence from prior participatory research and direct stakeholder evaluation. During the elicitation stage, we extracted guidelines derived from earlier participatory studies in the literature to ensure this knowledge was incorporated in our study without requiring us to repeat each of the participant interactions those studies involved. In the validation stage, prototype evaluation and stakeholder feedback were gathered by conducting interviews with community members and cognitive walkthroughs with personas representing the four vulnerable groups.

\section{Requirements Elicitation}
Following our human-centred requirements engineering framework, we first conducted a targeted, non-systematic review of both academic literature and grey literature (e.g., design guidelines from accessibility organisations, papers in archives, and blogs). The review drew from both HCI and software engineering literature, including foundational works on interface design \cite{norman2013design, shneiderman2021designing}, accessibility standards \cite{WebConte89_online}, empirical studies on specific vulnerable populations \cite{gomez2023design, srivastava2021actionable}, and software engineering research on accessibility barriers in mobile applications \cite{zaina2022preventing, tan_modified_2020}. This was a deliberate choice as inclusive design guidance specifically for safety-critical systems does not yet exist as an established body of work, and drawing on both HCI and SE literature ensured the guidelines were grounded in the strongest available evidence across both fields. One contribution of this study lies in translating these guidelines into implementable RE artefacts: functional and non-functional requirements with measurable criteria, traceable from elicitation through to validated prototype features. This approach prioritised breadth over exhaustiveness, aiming to capture guidelines from previous research and practice. A key feature of this stage was that many of the guidelines extracted from the literature were outcomes of earlier participatory studies, where researchers directly engaged with these communities through interviews, workshops, and field observations. With this, we ensured that the elicitation stage incorporated the lived experiences of vulnerable users without duplicating previous data-collection efforts.

\begin{enumerate}
    \item The extracted guidelines were grouped into our four user groups based on the original literature
    \item Examined the guidelines to identify potential overlaps and shared needs.\\
    This process revealed that while many guidelines were specific to a single group (e.g., high-contrast icons for colour-impaired users), a significant number were common to two or more groups. For instance, large, well-spaced buttons supported both elderly and people with low digital literacy, and a simplified navigation structure helped elderly, rural residents and individuals with low digital literacy as it simplified systems. 
    \item Present these guidelines as ``Guidelines specific to a single vulnerable community" [Table \ref{tab_guidelines}], and ``Guidelines Shared between Vulnerable Communities" [Table \ref{tab_shared_guidelines}]. These tables only show an extraction from the original guidelines, and the detailed guidelines can be found in Appendix A and B. 
\end{enumerate}

\subsection{Guidelines: Specific to a Single Vulnerable Community }
Table \ref{tab_guidelines} shows an extraction of the design guidelines targeted at individual vulnerable groups. The detailed guidelines can be found in Appendix A. The first group, Elderly communities, contain 11 guidelines (E1-11), Low digital communities have 15 guidelines (D1-15), Rural communities have a smaller set of guidelines (R1-5) and lastly color blind communities have 10 guidelines (C1-10). The guidelines for the elderly mostly focus on simplification and readability of the systems to reduce cognitive overload and to address their possible physical challenges (motor control issues, low vision issues). Low digital literacy community guidelines focus on clear information architecture, the use of plain language and consistency to overcome any possible confusion. For Rural residents, the emphasis is on the connectivity constraints and how best the systems can support them. For the colour blind communities guidelines, highlights the need for redundant cues beyond colour — e.g., patterns, icons, and text labels alongside colour coding. 

\begin{tabularx}{\linewidth}{>{\textwidth=0.12\textwidth}l >{\textwidth=0.88\textwidth}X}
\caption{Guidelines for Vulnerable Communities}\label{tab_guidelines} \\

\toprule
\textbf{ID} & \textbf{Guideline: Description} \\
\midrule
\endfirsthead

\toprule
\textbf{ID} & \textbf{Guideline: Description (continued)} \\
\midrule
\endhead

\midrule
\multicolumn{2}{r}{\textit{Continued on next page}} \\
\midrule
\endfoot

\bottomrule
\endlastfoot

\multicolumn{2}{l}{{\Large \faBlind}\, \textbf{Elderly Communities}} \\
\specialrule{1.2pt}{0pt}{0pt}

E1  & \textbf{Reduce short-term memory load:} Reducing the cognitive load is important for the elderly who may have decreased working memory and for those who may not be very familiar with digital systems. \cite{shneiderman2021designing}\\
E2  & \textbf{Maintain focus on current action:} Older users have more trouble concentrating. \cite{gomez2023design}\\
... & ... \\
E11 & \textbf{Prefer tapping over gestures:} Older users may have motor skill challenges that make gestures difficult, which is compounded by a lack of familiarity. \cite{gomez2023design}\\

\midrule
\multicolumn{2}{l}{{\Large \faKeyboard}\, \textbf{Low-Digital-Literacy Communities}} \\
\specialrule{1.2pt}{0pt}{0pt}

D1  & \textbf{Include short, simple instructions in Help menu:} Ensure help menus are short and easy-to-understand without jargon and are quick to action \cite{srivastava2021actionable}\\
D2  & \textbf{Enable customisation:} Allow users to customise the content, layout, and other settings for the application \cite{srivastava2021actionable}\\
... & ... \\
D15 & \textbf{Error-free operation:} Apps should be as error-free as possible and, when they fail, should fail “gracefully” instead of crashing \cite{tan_modified_2020}\\

\midrule
\multicolumn{2}{l}{{\Large \faTree}\, \textbf{Rural Communities}} \\
\specialrule{1.2pt}{0pt}{0pt}

R1  & \textbf{Split content into pages:} Pagination enables partial loading to help in areas with poor Internet access \cite{zaina2022preventing}\\
R2  & \textbf{Promote your app/website – make it easy to find:} Rural-facing content is easily overwhelmed in search results due to poor Search Engine Optimisation or being out-competed by generalised content \cite{groves2000web}\\
... & ... \\
R5  & \textbf{Rely on telecommunication rather than Internet:} Internet access can be poor, but other connectivity methods like direct sharing are often available \cite{roberts2017review}\\

\midrule
\multicolumn{2}{l}{{\Large \faLowVision}\, \textbf{Color-Blind Communities}} \\
\specialrule{1.2pt}{0pt}{0pt}

C1  & \textbf{Test against tools:} Use existing accessibility assessment tools such as WCAG tool suites \& colour-blindness simulators \cite{HowToDes35_online}\\
C2  & \textbf{Use more than just colour to distinguish things:} Don’t rely solely on colour to inform users; add other indicators like symbols or shapes \cite{ColorBli89_online}\\
... & ... \\
C10 & \textbf{Colour-blind users have different preferences:} Colours such as red and green have common meanings for most users, but colour-blind users may interpret them differently \cite{HowToDes35_online}\\
\end{tabularx}

\subsection{Guidelines: Shared between Vulnerable Communities}
In addition to group-specific guidance, we also identified a significant number of overlapping design guidelines across the four groups. During our analysis, where guidelines targeted similar usability or accessibility concerns, we classified them as shared guidelines. We found 3 guidelines (EC1-3) common to elderly and colour-blind groups, with the focus being on improving colour contrast for better readability. Between elderly and rural communities, there were 8 common guidelines (ER1-8), and for rural and low digital literacy communities, there were 5 guidelines (RD1-5). We also identified 5 guidelines (ERD1-5) that were common to 3 groups: elderly, rural and low digital literacy communities and these focused on providing more feedback and simplifying systems. An extraction of these guidelines are shown based on the shared grouping in table \ref{tab_shared_guidelines} below. The detailed guidelines can be found in Appendix B.

\begin{tabularx}{\linewidth}{>{\hsize=0.12\hsize}l >{\hsize=0.88\hsize}X}
\caption{Shared Guidelines Across Communities}\label{tab_shared_guidelines} \\

\toprule
\textbf{ID} & \textbf{Guideline: Description} \\
\midrule
\endfirsthead

\toprule
\textbf{ID} & \textbf{Guideline: Description (continued)} \\
\midrule
\endhead

\midrule
\multicolumn{2}{r}{\textit{Continued on next page}} \\
\midrule
\endfoot

\bottomrule
\endlastfoot

\multicolumn{2}{l}{{\Large \faBlind\ \ \faLowVision}\, \textbf{Elderly \& Color-Blind Communities}} \\
\specialrule{1.2pt}{0pt}{0pt}

EC1  & \textbf{Colour contrast between popups and background:} Ensure that colours used between foregrounds and backgrounds have high contrast to cater for potential vision challenges \cite{gilbert2019inclusive}\\
EC2  & \textbf{Design without colour to maintain luminance contrast:} If colours are used to code or contrast two elements, revise the design without colour (i.e., without chromaticity) to compare them in grayscale (luminance or brightness) \cite{Colorbli3_online}\\
EC3  & \textbf{Maintain WCAG contrast ratios:} Ideally, follow the AAA standard of 4.5:1 for text, as well as other ratios for different scenarios, to cater to low contrast vision and colour-blind users. Make use of compliance checkers (e.g., Google Lighthouse) to validate that designs meet required contrast ratios \cite{HowToDes35_online}\\

\midrule
\multicolumn{2}{l}{{\Large \faBlind\ \ \faTree}\, \textbf{Elderly \& Rural Communities}} \\
\specialrule{1.2pt}{0pt}{0pt}

ER1  & \textbf{Give time to act and read:} Avoid deadlines or timed events so people with different reading speeds are not rushed or cut off \cite{gomez2023design}\\
ER2  & \textbf{Discoverability – clear purpose and state:} Clear focal points, good visual hierarchy, and transparent navigation so users always know what something does and its current state \cite{norman2013design}\\
... & ... \\
ER8  & \textbf{Provide text alternatives:} Provide alternatives (e.g., alt-text, large text, or simplified language) for non-text content \cite{WebConte89_online}\\

\midrule
\multicolumn{2}{l}{{\Large \faTree\ \ \faKeyboard}\, \textbf{Rural \& Low-Digital-Literacy Communities}} \\
\specialrule{1.2pt}{0pt}{0pt}

RD1  & \textbf{Minimise download times:} Use smaller image/video sizes, compression, and substitute images with text where possible to improve performance on low bandwidth \cite{groves2000web}\\
RD2  & \textbf{Consider alternative information infrastructure:} Support cases where multiple users share one device and ensure offline distribution of the app if app stores are not accessible \cite{hardy2019designing}\\
... & ... \\
RD5  & \textbf{Enable offline access:} Ensure key features remain accessible offline when network coverage is poor or intermittent \cite{Whataret15_online}\\

\midrule
\multicolumn{2}{l}{{\Large \faBlind\ \ \faTree\ \ \faKeyboard}\, \textbf{Elderly, Rural \& Low-Digital-Literacy Communities}} \\
\specialrule{1.2pt}{0pt}{0pt}

ERD1 & \textbf{Simplify navigation structure:} Flatten menu hierarchies and prefer linear navigation to reduce cognitive load \cite{srivastava2021actionable}\\
ERD2 & \textbf{Reduce alternative paths:} Ensure navigation paths are simple and consistent; minimise branching or multiple routes that may confuse users \cite{gomez2023design}\\
... & ... \\
ERD5 & \textbf{Understandable error messages:} Provide clear, actionable error messages so users know what went wrong and how to recover \cite{gilbert2019inclusive}\\
\end{tabularx}

\section{Requirements Specification} \label{sec_reqSpec}
Following the elicitation stage, we systematically interpreted and transformed the extracted inclusive design guidelines into implementable software requirements using both our knowledge and information in the original research. The goal of this stage was to bridge the gap between conceptual guidance from literature and practical specifications that could be implemented during development. Each guideline was interpreted in terms of what the system must do (functional requirement) or how it must perform/behave (non-functional requirement).

\subsection{Translation Process}
We followed a structured process to convert each guideline into a precise requirement. This approach ensured traceability from original evidence-based guidelines to system-level requirements, while enabling later validation and visualisation with the prototype.
\begin{enumerate}
    \item Traceability mapping: every requirement was assigned an identifier (R\#) that directly links it back to the originating guideline (e.g., E1 → R01)
    \item Requirement type classification: requirements were categorised as Functional (E.g. "provide context-aware help") or Non-Functional (E.g. "achieve WCAG-AA colour contrast")
    \item Sub-categorisation of Non-Functional requirements: non-functional requirements were further tagged as Usability, Accessibility, Performance/Efficiency, or Reliability, reflecting the core quality attribute being addressed
\end{enumerate}

To support consistency in the translation process, we applied a set of decision rules. A guideline was translated into a functional requirement when it described a specific behaviour or action the system needed to perform (e.g., "provide a back button on every screen"). In contrast, a guideline was translated into a non-functional requirement when it described a quality constraint on how the system should behave (e.g., "maintain WCAG AA contrast ratios"). Non-functional requirements (NFRs) were then further classified by the main quality concern they addressed. Usability NFRs focused on learnability, efficiency, and cognitive load; accessibility NFRs focused on the needs of users with sensory or motor impairments; performance NFRs focused on speed and resource efficiency; and reliability requirements focused on system stability and graceful failure. Where a guideline contained multiple distinct concerns, we separated it into multiple atomic requirements so that each could be implemented and validated independently. This was done, for example, for guidelines E1, D2, ER2, and ERD1. The initial translation was carried out by the first author and then reviewed by the remaining authors. Any disagreements were resolved through discussion. 

Some non-functional requirements in this catalogue include measurable criteria or references to specific standards. For example, R47 specifies WCAG AA/AAA contrast ratios (e.g., 4.5:1 for text) and R62 specifies a maximum navigation hierarchy depth of two levels. These should not be interpreted as fixed design requirements. Rather, they reflect the need to make abstract quality goals concrete enough to be evaluated. This follows Chung et al. \cite{chung2012non}, who argue that non-functional requirements become more actionable when quality goals are operationalised into measurable criteria. It is also refleted in established accessibility standards such as WCAG 2.2 \cite{WebConte89_online}, which specify quantitative thresholds to enable compliance verification. In this catalogue, measurable criteria are included to support validation and verification, not to unnecessarily constrain implementation choices.

\subsection{Specification} \label{subsec_spec}

\begin{figure}[h]
\centering
\includegraphics[width=0.95\textwidth]{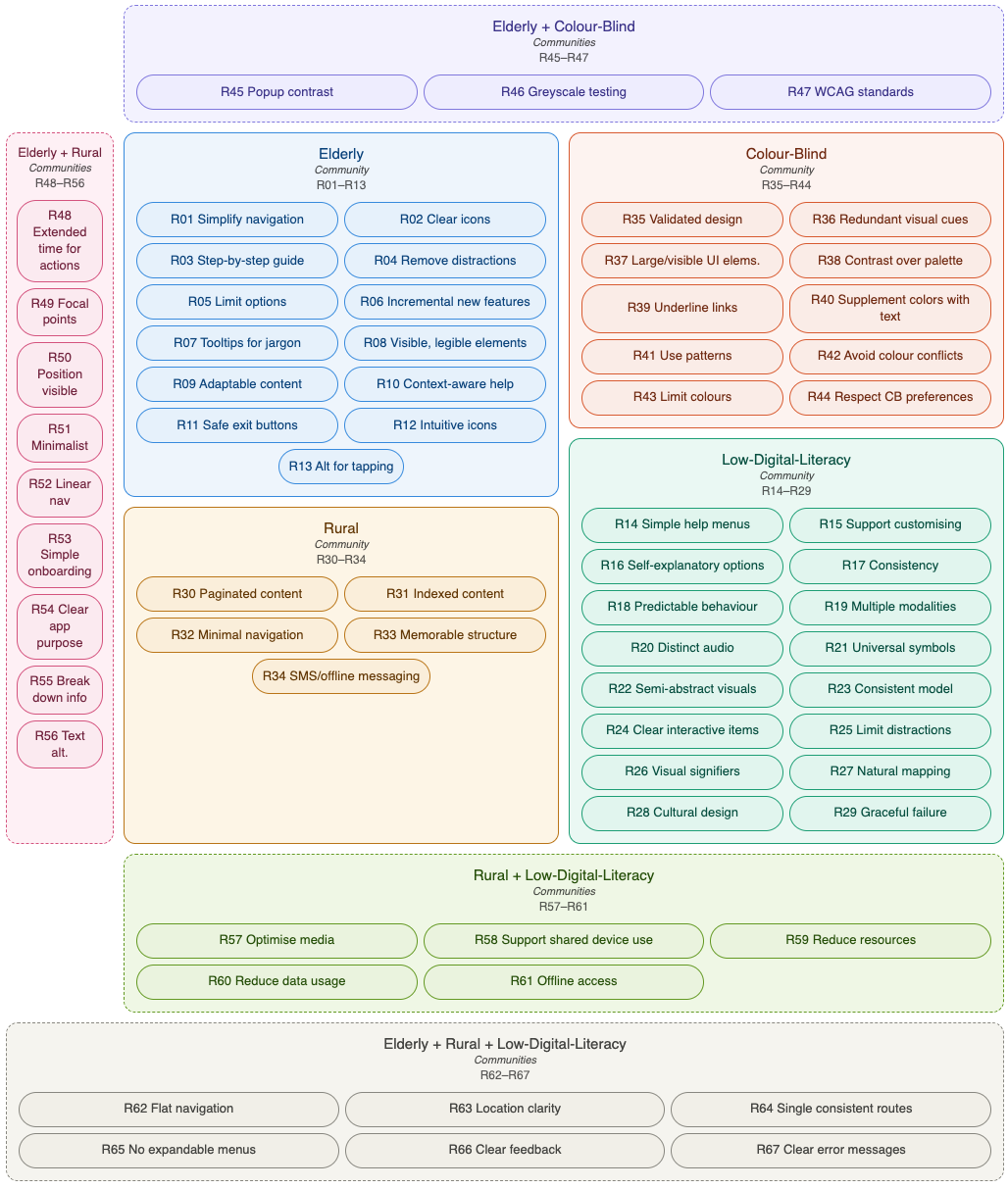}
\caption{Overview of Requirement Distribution}
\label{fig:req-grid}
\end{figure}

This specification consists of 67 requirements with 17 Functional requirements and 50 Non-Functional requirements. Of the non-functional requirements, 27 focused on usability, 19 on accessibility, 3 on performance/efficiency, and 1 on reliability. We present these requirements through four complementary artefacts. Table~\ref{tab_req_distribution} summarises the numerical distribution of requirements across community groups and non-functional categories. Figure~\ref{fig:req-grid} provides a holistic visual overview of how the requirements are distributed and shared across groups, showing the requirement labels spatially organised by community. Table~\ref{tab_guideline_requirements} presents an extraction of the full requirements to illustrate the level of detail and the guideline-to-requirement mapping. The complete specification, including all 67 requirements with their source guidelines and requirement types, is provided in Appendix C.

\begin{table}[H]
\centering
\caption{Distribution of Requirements by User Group and Non-Functional Categories (from final requirements table)}
\label{tab_req_distribution}
\begin{tabularx}{\textwidth}{lc|cc|*{4}{>{\centering\arraybackslash}X}}
\toprule
\textbf{User Group} & \textbf{Total Reqs.} & \textbf{Func.} & \textbf{Non-Func.} &
\multicolumn{4}{c}{\textbf{Non-Functional Categories}} \\
\cmidrule(lr){5-8}
 &  &  &  & \textbf{U} & \textbf{A} & \textbf{P} & \textbf{R} \\
\midrule
\faBlind\ Elderly                            & 13 & 3 & 10 & 7 & 3 & 0 & 0 \\
\faKeyboard\ Low-Digital-Literacy            & 16 & 5 & 11 & 8 & 3 & 0 & 0 \\
\faTree\ Rural                               & 5  & 3 & 2  & 2 & 0 & 0 & 0 \\
\faLowVision\ Colour-Blind                   & 10 & 0 & 10 & 1 & 9 & 0 & 0 \\
\midrule
\rowcolor{gray!15}
\multicolumn{8}{l}{\textbf{Shared Across Communities}} \\
\rowcolor{gray!10}
\quad \faBlind\ \faLowVision\ Elderly + Colour-Blind                          & 3  & 0 & 3  & 0 & 3 & 0 & 0 \\
\rowcolor{gray!10}
\quad \faBlind\ \faTree\ Elderly + Rural                                      & 9  & 3 & 6  & 5 & 1 & 0 & 0 \\
\rowcolor{gray!10}
\quad \faTree\ \faKeyboard\ Rural + Low-Digital-Literacy                      & 5  & 2 & 3  & 0 & 0 & 3 & 0 \\
\rowcolor{gray!10}
\quad \faBlind\ \faTree\ \faKeyboard\ Elderly + Rural + Low-Digital-Literacy  & 6  & 1 & 5  & 5 & 0 & 0 & 0 \\
\midrule
\rowcolor{gray!15}
\textbf{Shared Total}                & 23 & 6 & 17 & 10 & 4 & 3 & 0 \\
\midrule
\textbf{Grand Total}                 & \textbf{67} & \textbf{17} & \textbf{50} & \textbf{27} & \textbf{19} & \textbf{3} & \textbf{1} \\
\bottomrule
\multicolumn{8}{l}{\footnotesize U = Usability; A = Accessibility; P = Performance/Efficiency; R = Reliability.}
\end{tabularx}
\end{table}

Based on Table~\ref{tab_req_distribution}, the Elderly group shows a balance between functional and non-functional requirements (3 functional vs 10 non-functional), with usability concerns (7) dominating. The Low-Digital-Literacy group has a slightly higher share of functional requirements (5/16), yet still prioritises usability and cultural responsiveness. Rural users are associated with fewer requirements overall (5). In contrast, Colour-Blind users contribute mostly accessibility-focused non-functional requirements (9/10). The requirements shared between groups had more non-functional than functional requirements (17/6). 

Figure~\ref{fig:req-grid} provides a holistic view of how the 67 requirements are distributed across individual and shared community groups, illustrating how inclusive design needs overlap across vulnerable populations.

\newpage
An extraction of the detailed requirements for each of these groups is presented in Table \ref{tab_guideline_requirements}; the full list of requirements is in Appendix C. 

\begin{tabularx}{\textwidth}{p{0.28\textwidth} X p{0.15\textwidth}}
\caption{Guidelines and Corresponding Requirements with Types}\label{tab_guideline_requirements} \\

\toprule
\textbf{Guideline (ID: Name)} & \textbf{Requirement (R\#)} & \textbf{Req.\ Type} \\
\midrule
\endfirsthead

\toprule
\textbf{Guideline (ID: Name)} & \textbf{Requirement (R\#)} & \textbf{Req.\ Type} \\
\midrule
\endhead

\midrule
\multicolumn{3}{r}{\textit{Continued on next page}} \\
\midrule
\endfoot

\bottomrule
\endlastfoot

\multicolumn{3}{l}{{\Large \faBlind}\ \textbf{Elderly Communities}} \\
\specialrule{1.2pt}{0pt}{0pt}

\multirow{3}{*}{\textbf{E1:} Reduce short-term memory }
    & \textbf{R01:} Ensure the system simplifies navigation with a consistent and predictable structure.
    & Non-functional (Usability) \\

    load \cite{shneiderman2021designing} & \textbf{R02:} Ensure icons are clear, consistent, and semantically unambiguous across all screens.
    & Non-functional (Usability) \\

    & \textbf{R03:} Ensure the system provides step-by-step instructions for key tasks.
    & Non-functional (Usability) \\

\textbf{E2:} Maintain focus on current action \cite{gomez2023design}
    & \textbf{R04:} Ensure the interface removes distractions and secondary functions, highlighting the most important or typical action on each screen.
    & Non-functional (Usability) \\

... & ... & ... \\

\textbf{E11:} Prefer tapping over gestures \cite{gomez2023design}
    & \textbf{R13:} Ensure tapping alternatives exist for gesture-based interactions (e.g. sliders) to reduce dexterity demands on the user.
    & Functional \\

\specialrule{1.2pt}{0pt}{0pt}
\multicolumn{3}{l}{{\Large \faKeyboard}\ \textbf{Low-Digital-Literacy Communities}} \\
\specialrule{1.2pt}{0pt}{0pt}

\textbf{D1:} Include short, simple Help instructions \cite{srivastava2021actionable}
    & \textbf{R14:} Ensure help menus provide concise, jargon-free, and easy-to-act-on instructions.
    & Non-functional (Usability) \\

\multirow{2}{*}{\textbf{D2:} Enable customisation \cite{srivastava2021actionable}}
    & \textbf{R15:} Ensure users can customise content and layout via settings to fit their needs.
    & Functional \\

    & \textbf{R16:} Ensure customisation options are self-explanatory without requiring external help.
    & Non-functional (Usability) \\

... & ... & ... \\

\textbf{D15:} Error-free operation \cite{tan_modified_2020}
    & \textbf{R29:} Ensure the app is stable and fails gracefully with clear recovery options instead of crashing.
    & Non-functional (Reliability) \\

\specialrule{1.2pt}{0pt}{0pt}
\multicolumn{3}{l}{{\Large \faTree}\ \textbf{Rural Communities}} \\
\specialrule{1.2pt}{0pt}{0pt}

\textbf{R1:} Split content into pages \cite{zaina2022preventing}
    & \textbf{R30:} Ensure long content loads in smaller paginated chunks to improve usability under poor Internet conditions.
    & Functional \\

\textbf{R2:} Promote app/website visibility \cite{groves2000web}
    & \textbf{R31:} Ensure content is properly indexed (e.g., by search engines) and organised into discrete, discoverable pages.
    & Functional \\

... & ... & ... \\

\textbf{R5:} Rely on telecommunication \cite{roberts2017review}
    & \textbf{R34:} Ensure the app supports SMS or offline-friendly messaging for areas with poor Internet connectivity.
    & Functional \\

\specialrule{1.2pt}{0pt}{0pt}
\multicolumn{3}{l}{{\Large \faLowVision}\ \textbf{Colour-Blind Communities}} \\
\specialrule{1.2pt}{0pt}{0pt}

\textbf{C1:} Test against tools \cite{HowToDes35_online}
    & \textbf{R35:} Ensure the design is validated using WCAG and colour-blindness simulators.
    & Non-functional (Accessibility/Compliance) \\

\textbf{C2:} Use more than colour to distinguish \cite{ColorBli89_online}
    & \textbf{R36:} Ensure redundant visual cues (e.g., icons, text, patterns) supplement colour-coded information.
    & Non-functional (Accessibility) \\

... & ... & ... \\

\textbf{C10:} Respect colour-blind preferences \cite{HowToDes35_online}
    & \textbf{R44:} Ensure designs respect typical user preferences (e.g., favouring blue for clarity) and avoid negative shifts in meaning.
    & Non-functional (Accessibility) \\

\specialrule{1.2pt}{0pt}{0pt}
\multicolumn{3}{l}{{\Large \faBlind\ \ \faLowVision}\, \textbf{Elderly + Colour-Blind Communities}} \\
\specialrule{1.2pt}{0pt}{0pt}

\textbf{EC1:} Colour contrast between popups and background \cite{gilbert2019inclusive}
    & \textbf{R45:} Ensure popup backgrounds and foreground text meet high contrast ratios for readability.
    & Non-functional (Accessibility) \\

\textbf{EC2:} Design without colour to maintain luminance contrast \cite{Colorbli3_online}
    & \textbf{R46:} Ensure designs work in greyscale by testing contrast without chromatic elements.
    & Non-functional (Accessibility) \\

\textbf{EC3:} Maintain WCAG contrast ratios \cite{HowToDes35_online}
    & \textbf{R47:} Ensure the design complies with WCAG AA/AAA standards (e.g., 4.5:1 for text) and validates dynamically with tools like Google Lighthouse.
    & Non-functional (Accessibility/Compliance) \\

\specialrule{1.2pt}{0pt}{0pt}
\multicolumn{3}{l}{{\Large \faBlind\ \ \faTree}\, \textbf{Elderly + Rural Communities}} \\
\specialrule{1.2pt}{0pt}{0pt}

\textbf{ER1:} Give time to act and read \cite{gomez2023design}
    & \textbf{R48:} Ensure time-sensitive actions allow extended durations so slower readers are not disadvantaged.
    & Functional \\

\multirow{2}{*}{\textbf{ER2:} Improve discoverability \cite{norman2013design}}
    & \textbf{R49:} Ensure screens provide clear focal points and consistent visual hierarchy.
    & Non-functional (Usability) \\

    & \textbf{R50:} Ensure the interface always makes the user's current position and available return paths clearly visible.
    & Non-functional (Usability) \\

... & ... & ... \\

\textbf{ER8:} Provide text alternatives \cite{WebConte89_online}
    & \textbf{R56:} Ensure all non-text content includes descriptive text alternatives (e.g., alt-text).
    & Non-functional (Accessibility) \\

\specialrule{1.2pt}{0pt}{0pt}
\multicolumn{3}{l}{{\Large \faTree\ \ \faKeyboard}\, \textbf{Rural + Low-Digital-Literacy Communities}} \\
\specialrule{1.2pt}{0pt}{0pt}

\textbf{RD1:} Minimise download times \cite{groves2000web}
    & \textbf{R57:} Ensure multimedia elements are optimised or replaced with lightweight alternatives to improve load times on poor connections.
    & Non-functional (Performance) \\

\textbf{RD2:} Consider alternative information infrastructure \cite{hardy2019designing}
    & \textbf{R58:} Ensure app design supports shared-device usage and alternative distribution channels (e.g. share app package) when/if app stores are inaccessible.
    & Functional \\

... & ... & ... \\

\textbf{RD5:} Enable offline access \cite{Whataret15_online}
    & \textbf{R61:} Ensure essential features remain accessible offline or with intermittent connectivity.
    & Functional \\

\specialrule{1.2pt}{0pt}{0pt}
\multicolumn{3}{l}{{\Large \faBlind\ \ \faTree\ \ \faKeyboard}\, \textbf{Elderly + Rural + Low-Digital-Literacy Communities}} \\
\specialrule{1.2pt}{0pt}{0pt}

\multirow{2}{*}{\textbf{ERD1:} Simplify navigation }
    & \textbf{R62:} Ensure the app-level navigation hierarchy is flat, with no more than two levels of depth.
    & Non-functional (Usability) \\

    structure \cite{srivastava2021actionable} & \textbf{R63:} Ensure users can always determine their location within the app and return to a previous screen without confusion.
    & Non-functional (Usability) \\

\textbf{ERD2:} Reduce alternative paths \cite{gomez2023design}
    & \textbf{R64:} Ensure single, consistent routes for completing tasks to minimise confusion.
    & Non-functional (Usability) \\

... & ... & ... \\

\textbf{ERD5:} Understandable error messages \cite{gilbert2019inclusive}
    & \textbf{R67:} Ensure error messages are concise, instructive, and reduce user anxiety during recovery.
    & Functional \\

\end{tabularx}

\section{Requirements Prototyping}
As the next step of our human-centric requirements engineering for critical systems process, we focused on operationalising the specified requirements by designing a prototype of an Early Warning System (EWS) mobile application. There were two main reasons for choosing EWS as the subject of the prototype.
\begin{enumerate}
    \item \textbf{EWS platforms are archetypal safety-critical systems:} In EWS, delays, inaccessibility, or miscommunication can result in severe consequences, ultimately even leading to loss of lives. This domain provided an appropriate context to model and evaluate our human-centric requirements. It also helped to explore our suggested    
    socially responsible requirements engineering (RE) process, given its reliance on reliable information delivery and inclusive user interfaces.
    \item \textbf{Global need for more inclusive EWS:} EWS plays a key role in disaster communication and has functions like conveying disaster alerts, helping prepare emergency plans, and providing education on disaster management. The United Nations has recognised the importance of ensuring equal access in EWS via its initiative “Early Warnings for All,” which aims to ensure that every individual globally has access to EWSs by 2027. 
\end{enumerate}

In modelling the specified requirements via this prototype, the first step was to develop a core system prototype that incorporated the essential functions of a standard EWS \cite{syukron2024comprehensive}, such as alert notifications, event maps, and preparedness information. Building on this baseline, adaptive versions of the app were then developed for each user group—elderly, low-digital-literacy, rural, and colour-blind communities. These versions integrated the requirements specific to each vulnerable group while maintaining a consistent core architecture.

\subsection{Core System Prototype}
EWS can focus on a single or multiple hazards, such as floods, earthquakes, and bushfires. For our prototype, we focused on bushfires as this was the disaster that was most relevant to our group of researchers. 

The key functionalities of this prototype included 1) real-time, location-based notifications and news, 2) a map displaying nearby fires, 3) disaster preparation guides in various stages of the disaster: Plan and prepare, Get Ready, Take Action, Evacuate, and 4) contacting emergency services for assistance. 

\begin{figure}[h]
  \centering
  \setlength{\tabcolsep}{2pt}
  \begin{subfigure}{0.25\textwidth}\centering
    \includegraphics[width=\linewidth]{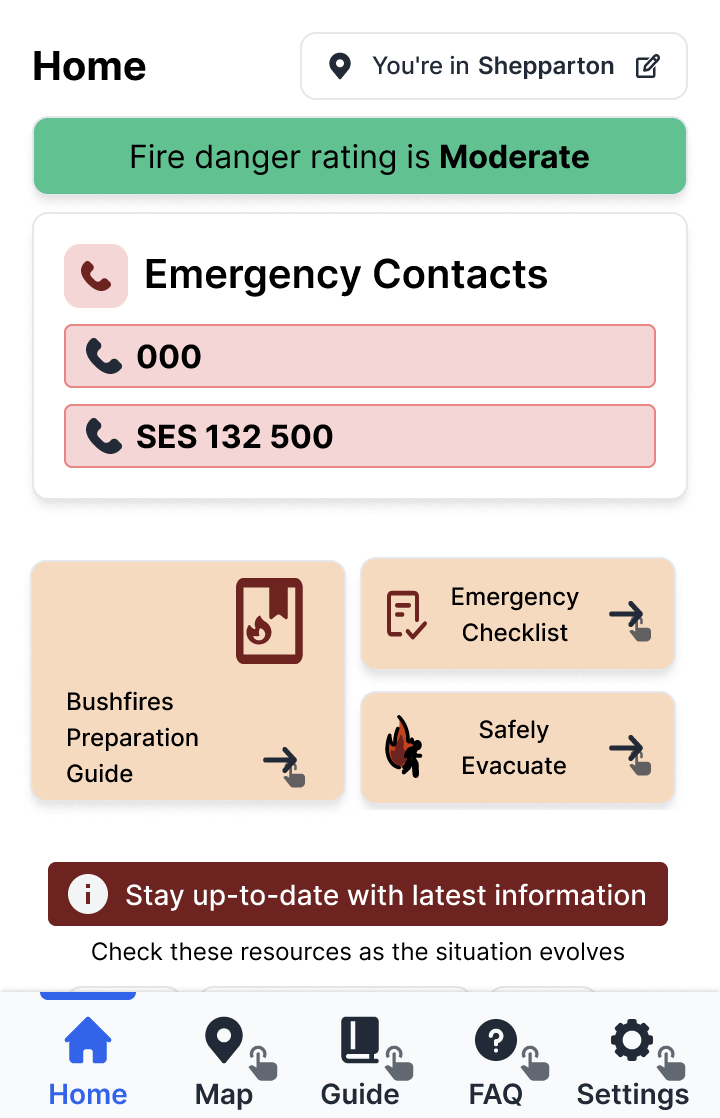}
    \caption{Home Page}
  \end{subfigure}
  \begin{subfigure}{0.25\textwidth}\centering
    \includegraphics[width=\linewidth,trim={0cm 0cm 0.3cm 0cm},clip]{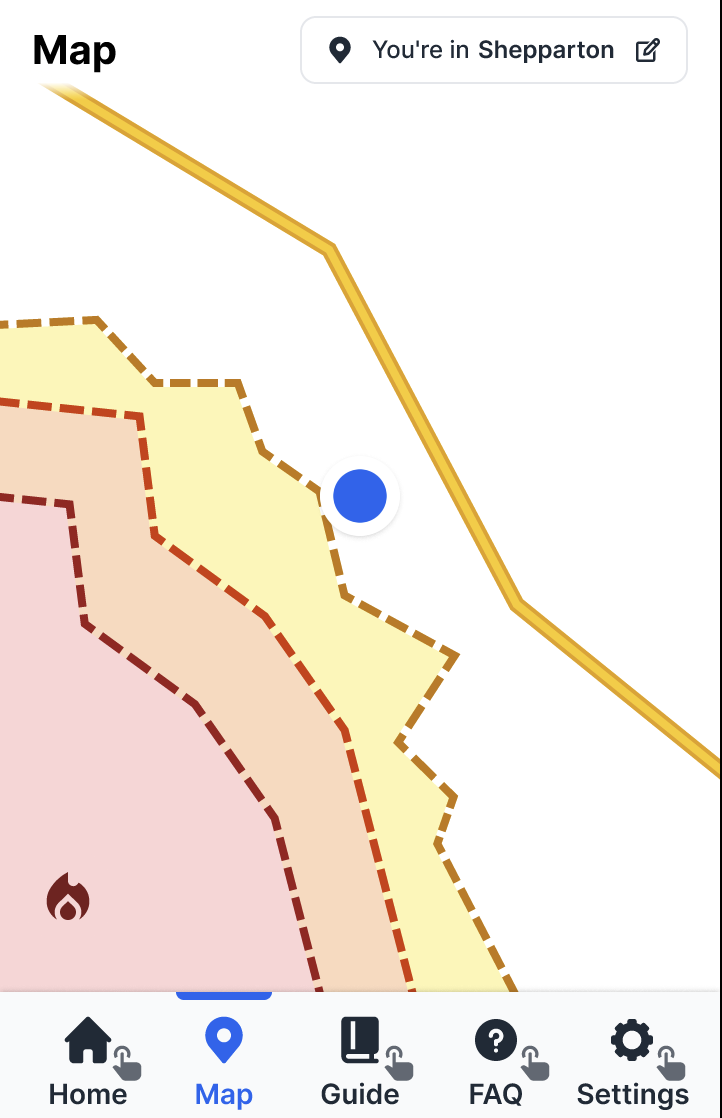}
    \caption{Map Page}
  \end{subfigure}
  \begin{subfigure}{0.25\textwidth}\centering
    \includegraphics[width=\linewidth,trim={0cm 0cm 0.3cm 0cm},clip]{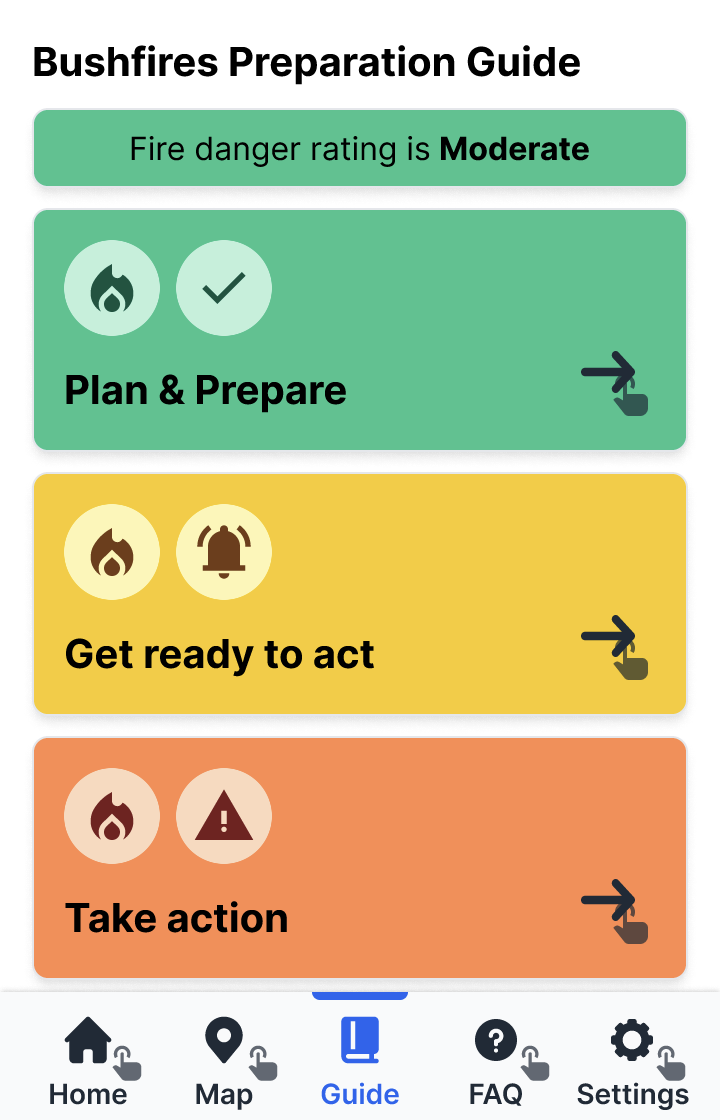}
    \caption{Guides Page}
  \end{subfigure}
  \caption{Core Prototype}
  \label{fig_corePrototype}
\end{figure}

As shown in Figure \ref{fig_corePrototype}(a), on the home page, users had immediate access to emergency services through a speed dial. There were quick links to critical content like the disaster (bushfire) preparation guide, emergency checklists and evacuation guide within the app. Additionally, location-based information retrieval was also integrated, displaying trusted information sources relevant to the user’s current location, e.g. for a user living in Victoria (Australia), it would display information from VicEmergency. It also had a live banner to show real-time updates and warnings e.g Fire danger rating is moderate. Additionally, the bottom navigation bar facilitated easy switching between different sections such as home, maps, guide, FAQ and settings.

In the map page (Figure \ref{fig_corePrototype}(b)), the current location of the user is displayed with a blue dot, and they will be able to see the current bushfire danger rating for their current location, and how far they are from the bushfire zones. This would be able to give them an estimation of the distance to danger and the ability to visualise the evacuation routes to avoid running towards fire. Most of the current disaster apps make it very difficult to determine evacuation routes based on the current status of a disaster. Therefore, we added this feature and focused on simplifying the information visualisation. In the final system design, this map page would ideally incorporate road networks and would pull data from a source such as Google Maps. 

The disaster preparation guide (Figure \ref{fig_corePrototype}(c)), a FAQ page with some common questions and contact details for the developer team and a settings page which allowed customisation of the app. The customisations supported in the settings page will be explained in detail in the next section. 

\begin{figure}[h]
  \centering
  \setlength{\tabcolsep}{2pt}
  \begin{subfigure}{0.24\textwidth}\centering
    \includegraphics[height=10cm,trim={0.6cm 1cm 50cm 0.5cm},clip]{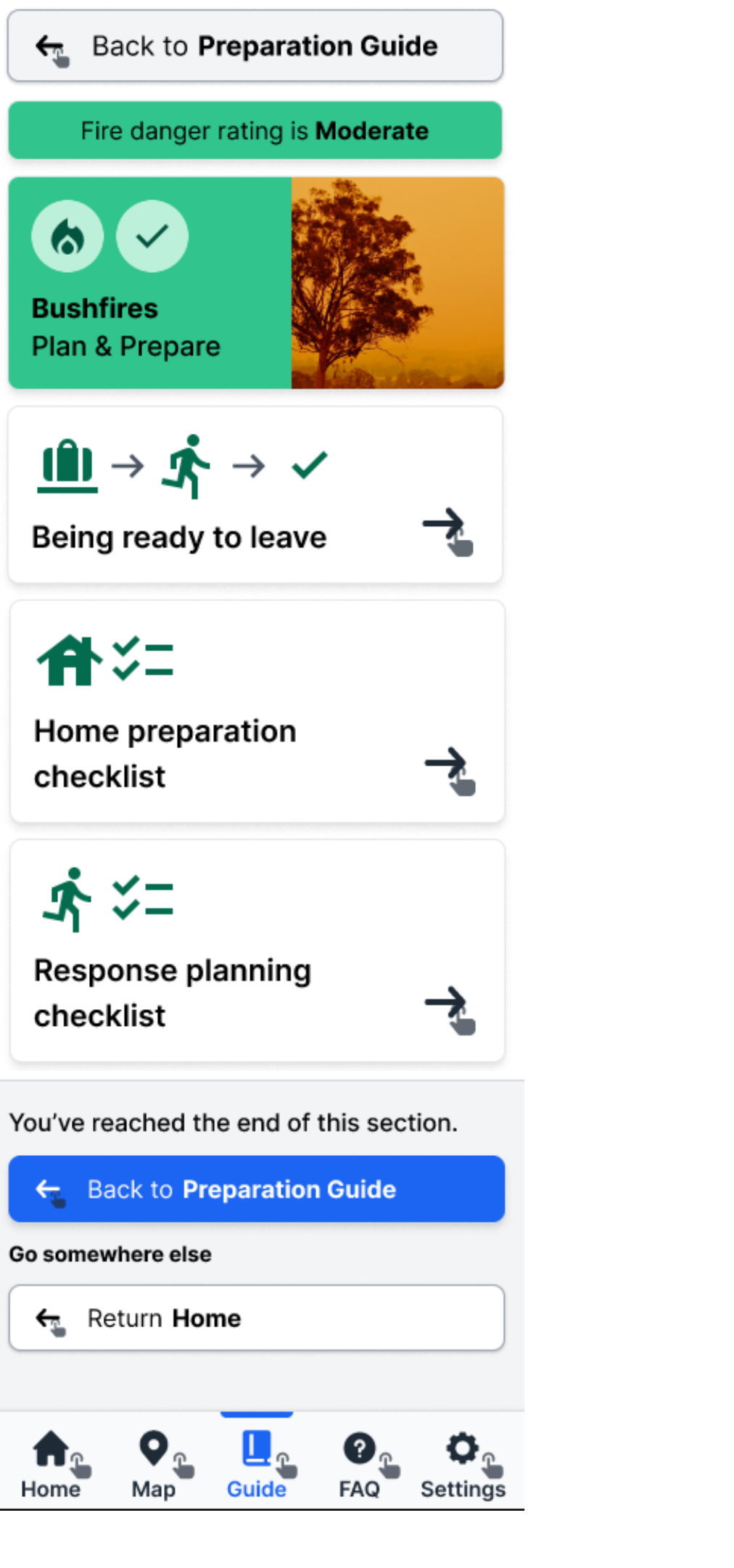}
    \caption{Plan \& Prepare Guide}
  \end{subfigure}
  \begin{subfigure}{0.24\textwidth}\centering
    \includegraphics[height=10cm,trim={0.8cm 1cm 50cm 0.5cm},clip]{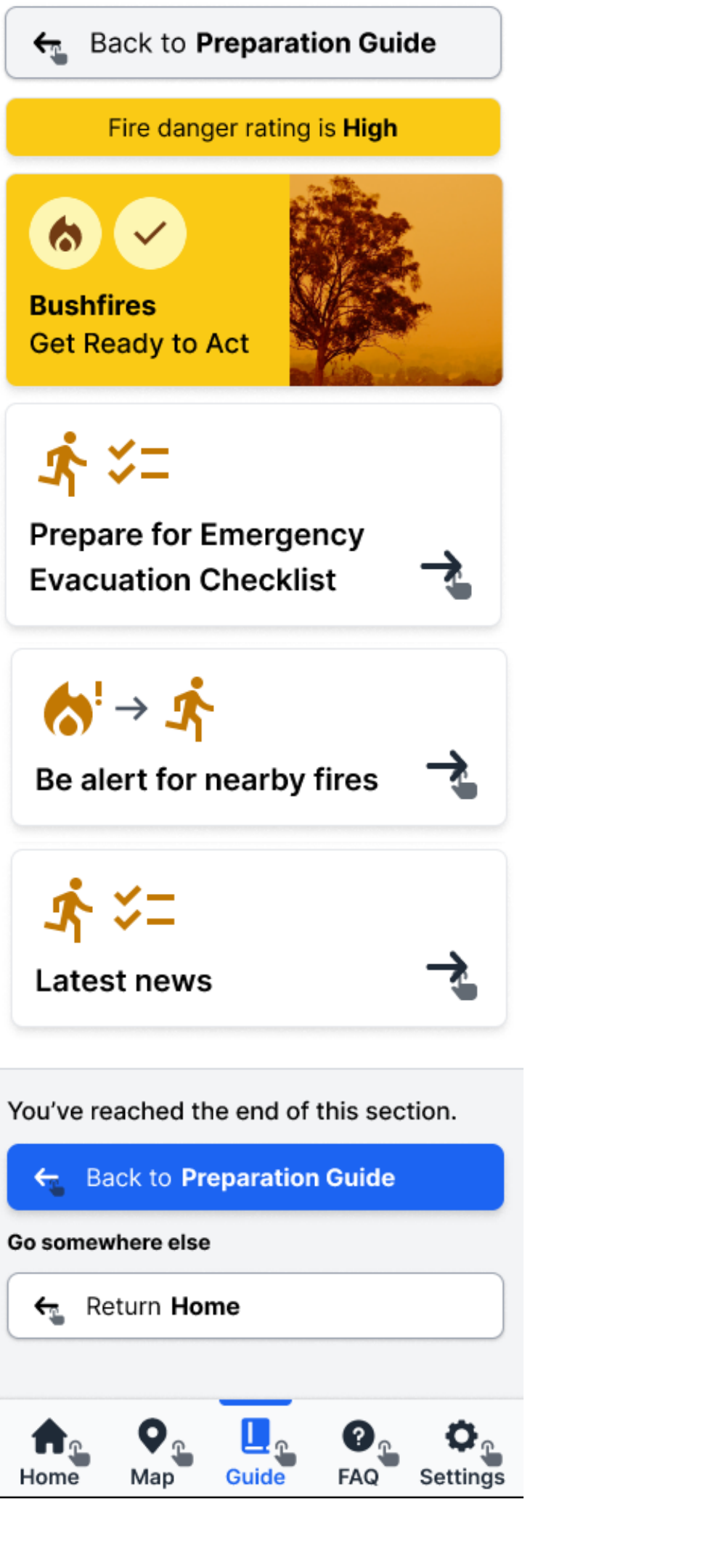}
    \caption{Ready to Act Guide}
  \end{subfigure}
  \begin{subfigure}{0.24\textwidth}\centering
    \includegraphics[height=10cm,trim={0.8cm 1cm 50cm 0.5cm},clip]{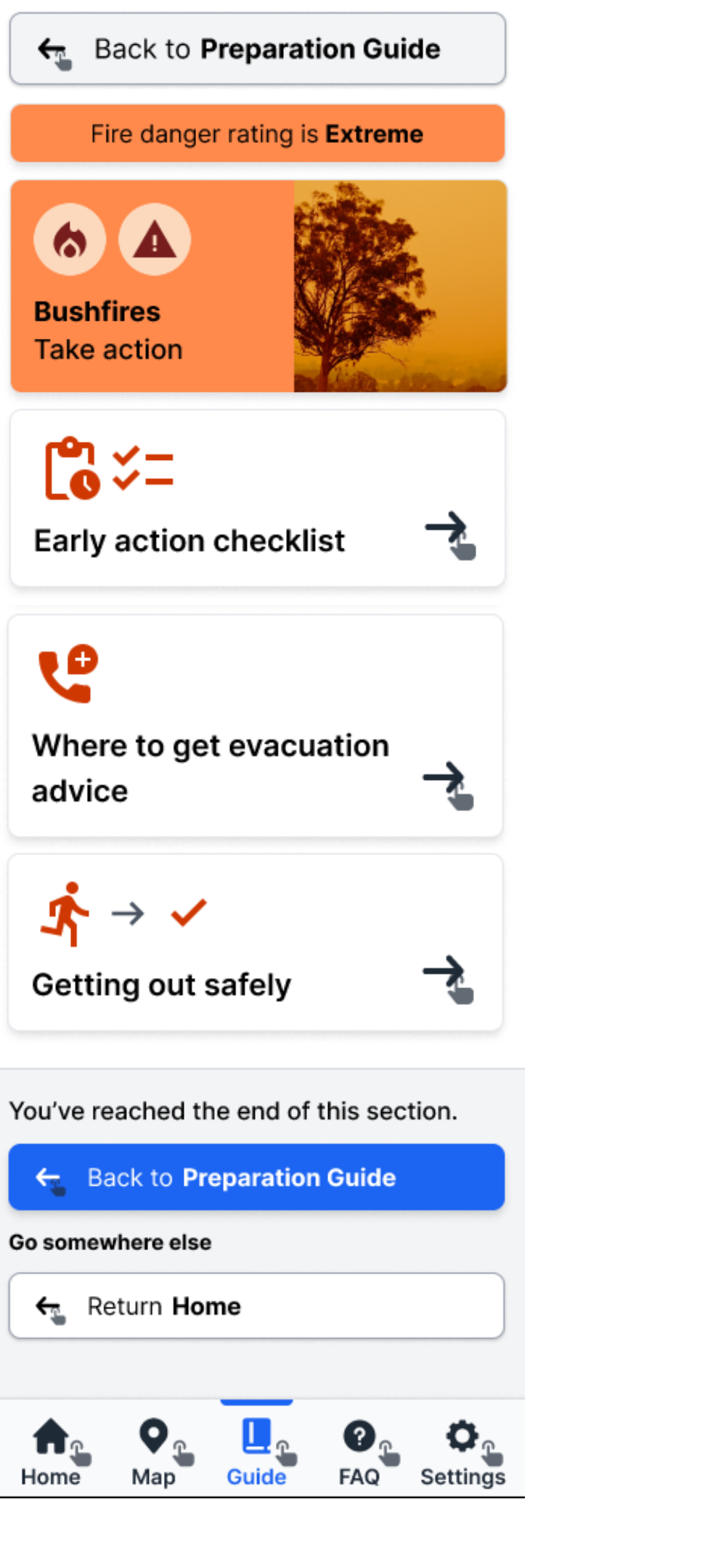}
    \caption{Take Action Guide}
  \end{subfigure}
  \begin{subfigure}{0.24\textwidth}\centering
    \includegraphics[height=10cm,trim={0.8cm 0.2cm 50cm 0.2cm},clip]{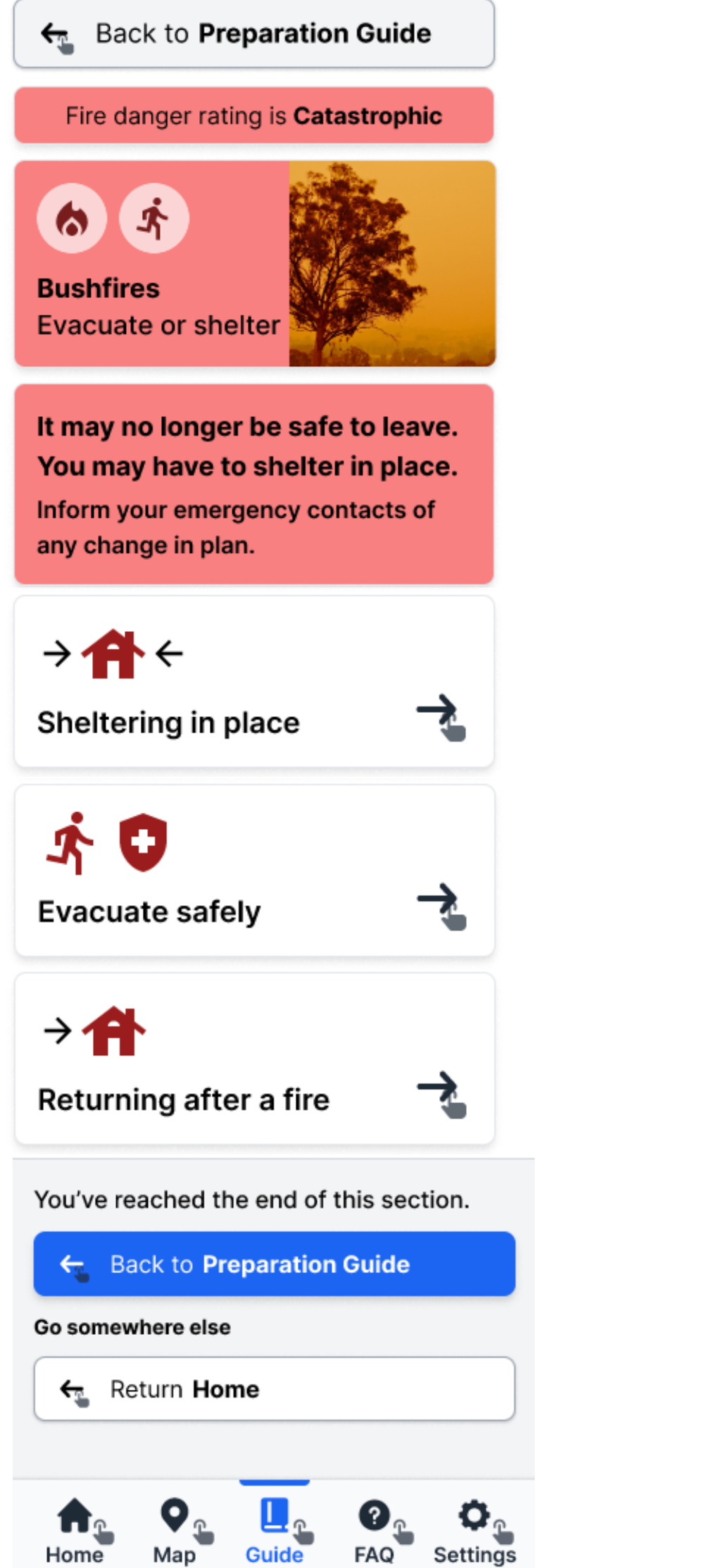}
    \caption{Evacuate or Shelter Guide}
  \end{subfigure}
  \caption{Guide Pages}
  \label{fig_guides}
\end{figure}

The disaster preparation guide was categorised into four: Plan and Prepare, Get Ready to Act, Take Action, Evacuate, and Shelter. Each guide was associated with a fire danger rating ranging from Moderate to Catastrophic, which are standard danger ratings published by the Country Fire Authority (CFA), Australia. We used colors from CFA fire danger ratings as the theme colour for each of the guides, to help users easily distinguish the relevant guides. Information for these guides was sourced from trusted bushfire management agencies such as Australian national and state governments \cite{bushfires_smarttraveller, bushfire_recovery, nsw_emergency_kite, abc_emergency_kit, qld_emergency_kit}. The list of prototype functions included in each of these guides is listed below. 

\begin{enumerate}
    \item \textbf{1. Plan and Prepare Guide:} Provides information on general disaster preparation [Moderate/No Fire Danger] (Figure \ref{fig_guides}(a))
    \begin{enumerate}
        \item Information on how to be ready to evacuate 
        \item A checklist to prepare their home for a bushfire disaster (e.g., cleaning up leaf litter) 
        \item A checklist to help plan the user’s response in case of a bushfire
    \end{enumerate}
    \item \textbf{2. Get Ready to Act Guide:} Provides information on preparing for imminent bushfires [High Fire Danger] (Figure \ref{fig_guides}(b))
    \begin{enumerate}
        \item A checklist to prepare for an emergency evacuation
        \item Information on nearby emergency shelters and routes through Google Maps service to these places
        \item Next four-day fire danger forecast for users' current area so that they could stay vigilant \cite{cfa_2023} 
        \item Information on historical bushfire data in the neighbourhood (data extracted from a database created by the government \cite{vicplan})
    \end{enumerate}
    \item \textbf{3. Take Action Guide: } Provides information on preparing to evacuate [Extreme Fire Danger] (Figure \ref{fig_guides}(c))
    \begin{enumerate}
        \item A checklist of early actions E.g prepare an emergency kit, decide on a safe meeting place for your family
        \item Where to get evacuation advice E.g Radio, links to trusted sources
        \item Advice on how to get out safely
    \end{enumerate}
    \item \textbf{4. Evacuate or Shelter Guide:} Provides information on evacuating [Catastrophic Fire Danger] (Figure \ref{fig_guides}(d))
    \begin{enumerate}
        \item An evacuation checklist: A streamlined checklist of essential items to minimise preparation time before evacuation \cite{bushfire_safety_guide} 
        \item Guide on how to shelter in their home
        \item How to evacuate safely 
        \item Post-fire safety tips to protect health upon returning home \cite{Helpafte99_online}
    \end{enumerate}
\end{enumerate}

\subsection{Adaptive Versions}
Although the requirements in Section~\ref{sec_reqSpec} were derived and organised by community groups, to ensure that the specific needs of each vulnerable population were captured, the prototype was deliberately designed to avoid enforcing these groupings on users. All requirements across the four community groups were incorporated into the system design; the self-customisation approach simply allowed users to activate the features relevant to them, rather than being assigned to a pre-defined group-based version. This was a deliberate design decision grounded in the intersectional nature of vulnerability. Individuals rarely belong to a single vulnerability category e.g an older adult may also have colour-vision impairment and live in a rural area. Therefore assigning users to fixed group-based versions would fail to reflect this reality. Instead, the prototype supported self-customisation, allowing users to select the accessibility and usability features most relevant to their own circumstances, regardless of which community group they belonged to. This approach preserved the traceability between requirements and prototype features while respecting the diversity and autonomy of individual users. Users could adjust visual, interaction, and content settings directly within the app to match their preferences and accessibility needs, including language, font size, location, ease-of-use settings, and colour scale for fire danger ratings. This ensured that the adaptation remained user-driven rather than system-imposed, giving users control over which accessibility and usability features mattered most to them.

\begin{figure}[h]
    \centering
\includegraphics[width=\columnwidth]{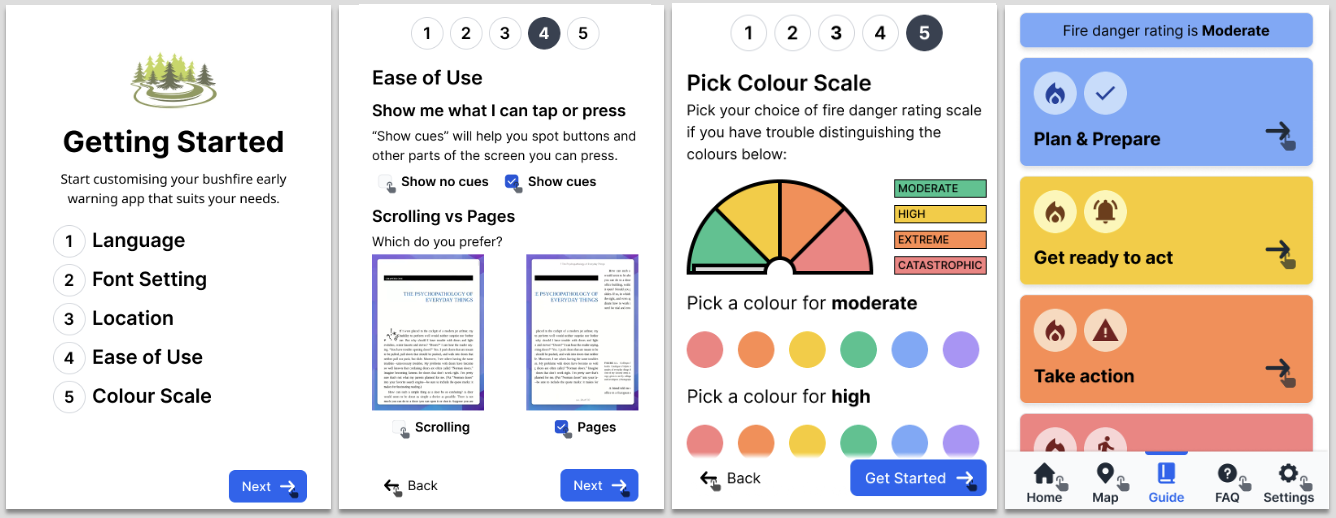}
    \caption{(a)-(c): Adapting the System at Set-up, (d): The adapted system}
    \label{fig:p_profiling}
\end{figure}

Users were able to define the adaptation parameters at the initial setup of the app, as shown in Figure \ref{fig:p_profiling}, and modify them later via the settings page.  We supported adapting the app language, font size, and location, ensuring that the alerts and interface were relevant and legible to the users. The app also allowed users to adjust its ease-of-use settings, such as choosing whether to display visual cues or selecting between scroll-based and page-based navigation depending on their comfort and familiarity. For visual accessibility, users could modify the colour scale used for fire danger ratings, selecting combinations that best suited their vision or colour-perception abilities. As these colours were also linked to the visual themes used in fire-danger warnings and preparedness guides, the colour-customisation feature played a crucial role in ensuring that critical safety information was clearly conveyed to every user. 

These options were included to reflect the diversity of user capabilities and contexts, while maintaining a single, unified system. Instead of relying on rigid, pre-defined versions for specific groups, this approach allowed individual users to decide which accessibility and usability features mattered most to them, creating a more flexible and inclusive early-warning experience.

\section{Prototype Evaluation and Stakeholder Feedback}

Following the specification and prototyping phases, we validated the identified requirements through participatory evaluation activities. These validations helped to examine how well the identified requirements were reflected in the prototype and to gather feedback for further refinement. In following this work's inclusive and human-centred nature, the validation process actively engaged end users and realistic user representations. We used two participatory approaches: semi-structured interviews with community members and cognitive walkthroughs using personas.

\subsection{Interviews}
We conducted prototype evaluation sessions with our target groups, combining structured task-based interaction with the prototype and semi-structured interview discussion. The task-based component assessed usability across the different user groups, while the interview discussion allowed participants to reflect on their experiences and provide open-ended feedback. Both components were used to generate the findings reported in Section~\ref{subsec_results}. Each interview took approximately 30-45 minutes. We advertised our study via social media platforms and personal contacts, which was a pragmatic recruitment approach given the time and resource constraints of the study. Participants were eligible if they self-identified as belonging to one of the target groups (older adults or rural residents), owned a smartphone, and were English-speaking. We received six responses and conducted Zoom-based online interviews with them.

\noindent
\captionof{table}{Prototype Evaluation Tasks: Interview}
\label{tab:evaluation-general-actions}
\begin{tabularx}{\textwidth}{l X}
\toprule
\textbf{ID} & \textbf{Task} \\
\midrule
1 & Open the app and navigate to the home screen. \\
2 & Access the emergency kit preparation checklist and tick some items. \\
3 & Locate and access the Settings. \\
4 & Access the map to see nearby disasters. \\
5 & Access the preparation guide for each of the bushfire danger ratings. \\
6 & Attempt to navigate back to the home page. \\
7 & Evaluate the overall user experience and ease of use. \\
\bottomrule
\end{tabularx}

During the interviews, participants were first introduced to the prototype and were asked to explore its core features, including the home, map, guides, FAQ, and settings pages. They were then tasked with completing seven predefined tasks as outlined in Table \ref{tab:evaluation-general-actions} by interacting with our prototype tool. These tasks were designed to test the prototype’s usability across the different user groups effectively. After the user performed these tasks, we conducted a discussion with them to understand their experience in using the prototype. We audio-recorded each interview, transcribed it later using Zoom's transcription feature and then performed thematic analysis on these transcriptions.

\subsection{Cognitive Walkthrough}
Due to difficulties in finding a sufficient number of participants to represent all four vulnerable user groups in our interviews, we employed cognitive walkthroughs as a complementary validation approach. As this standard validation method is widely used in HCI and software engineering when direct end-user testing is limited or not feasible, this suited our needs \cite{burnett2016gendermag,spencer2000streamlined}. This allowed us to systematically assess how users with varying characteristics, such as the elderly, rural, low digital literacy, and colour-impaired, would interact with the prototype.

\noindent
\captionof{table}{Tasks Used in Cognitive Walkthroughs}
\label{tab:cw_tasks}
\begin{tabularx}{\textwidth}{C|>{\raggedright\arraybackslash}p{0.05\textwidth}|X}
\toprule
\textbf{User Group} & \textbf{Task ID} & \textbf{Task Description} \\
\midrule

\multirow{4}{*}{\makecell{\Large \faBlind \\ Elderly}} 
& 1--7 & General tasks from Table \ref{tab:evaluation-general-actions}. \\
& 8 & Verify if font size adjustments adequately address readability needs. \\
& 9 & Evaluate the simplicity and clarity of instructions provided. \\
& 10 & Check if the interface design is straightforward and easy to navigate, considering a preference for simplicity. \\
\midrule

\multirow{4}{*}{\makecell{\Large \faKeyboard \\ Low-Digital\\Literacy}} 
& 1--7 & General tasks from Table \ref{tab:evaluation-general-actions}. \\
& 8 & Evaluate the simplicity and intuitiveness of navigation options. \\
& 9 & Verify that the app provides clear and concise guidance for adjusting text settings. \\
& 10 & Test the clarity and usefulness of step-by-step instructions, ensuring minimal technical jargon. \\
\midrule

\multirow{4}{*}{\makecell{\Large \faTree \\ Rural}} 
& 1--7 & General tasks from Table \ref{tab:evaluation-general-actions}. \\
& 8 & Evaluate the clarity and accessibility of the evacuation guides and map-based information. \\
& 9 & Verify that relevant, localised details about disaster and evacuation routes are provided \\
& 10 & Assess the usefulness of the app to support the safety of both family and livestock \\
\midrule

\multirow{4}{*}{\makecell{\Large \faLowVision \\ Colour-Blind}} 
& 1--7 & General tasks from Table \ref{tab:evaluation-general-actions}. \\
& 8 & Test the effectiveness of colour customisations in distinguishing interface elements. \\
& 9 & Evaluate alternative visual cues (e.g., patterns or text labels) for differentiating colours. \\
& 10 & Verify that critical information is not solely conveyed through colour. \\

\bottomrule
\end{tabularx}

We developed four personas to represent our four vulnerable communities: elderly, rural, low digital literacy, and colour-impaired. These personas are in Appendix D. A cohort of eight participants was recruited through personal networks to participate in the cognitive walkthrough sessions, with two assigned to each persona. At the beginning of each session, participants were first briefed on the characteristics, goals, and limitations of their assigned persona. For participants assigned to the colour-blind persona, we recommended using a colour-blind simulator to conduct the cognitive walkthroughs. For this group, a short demonstration was conducted on using a colour-blindness simulator to help them become familiar with it. Each participant then completed ten tasks during the walkthroughs. Seven of these tasks overlapped with those used in the interview study, while three were additional, task-specific to each user group. These tasks are listed in Table \ref{tab:cw_tasks}. After completing the walkthrough, participants were engaged in a discussion to reflect on their experiences. Each session lasted approximately one hour and was conducted online via Zoom. Similar to interviews, we audio-recorded each interview, transcribed it later using Zoom's transcription feature, and then performed thematic analysis of the transcriptions.

\subsection{Results} \label{subsec_results}
In the interview study, we had six participants: two elderly (age> 70) and four from rural areas. For the Cognitive walkthrough, we had eight participants, with two per persona. Upon performing thematic analysis of the discussions, we found that the feedback from both cohorts can be categorised into four main themes, which are discussed in detail in the next subsections. 

\subsubsection{Functionality}
The majority of participants in both interviews [INT] and cognitive walkthroughs [CW] found the functions presented in the prototype useful and clear. They found that the app's checklists were helpful for preparing for a disaster. One rural resident[INT] who works in emergency management observed that the app had ``some really good [..] equipment/views and other things to remind people [on what they needed to pack]". Several rural residents[INT] also appreciated the home page's quick links to guide pages, which made the most important instructions immediately accessible when the app is opened. They believed this quick access to information would be especially helpful when they are in the middle of a disaster situation, E.g Fire danger rating being Extreme and needing to be ready to evacuate. However, external links to other resources had divided opinions. While some rural users[INT] appreciated direct links to Google Maps for navigation, elderly users [INT] struggled to return to the app after following these links, often requiring assistance to reopen the prototype. In terms of issues in the app, some users were unsure whether any action was needed once a danger rating appeared [CW]; others did not recognise that the blue dot on the map represented their current location [INT]. There was also confusion caused by inconsistent text, such as ``You won’t be able to get emergency assistance,” in the Catastrophic Fire danger guide, which conflicted with the availability of a contact button[CW]. 

Participants also provided suggestions on improving the app. Some requested more detailed evacuation route information (e.g a map with wind information to help determine evacuation route)[CW] and quick links to safe points and local emergency services[CW]. The absence of such links was seen as a gap in supporting timely decision-making during disasters. Some participants also identified the need for stronger error-prevention mechanisms. This was highlighted by participants [CW] who were concerned about accidentally tapping on emergency contacts without confirmation, which could lead to unintended actions during stressful situations, and by an elderly participant [CW] who suggested restricting users from assigning the same colour to multiple danger ratings. They also suggested adding functions to automatically notify emergency services when the user is in danger [CW], to send reminders for unchecked checklist items [CW], and to maintain a log of recent alerts [CW]. A participant representing a low-digital-literacy persona also suggested allowing the users to personalise content E.g. adding new items to checklists and changing checklist item order. Feedback such as “it would be good to have an option at the beginning asking if notifications are required”[CW] and ``a button to call for help straightaway from the fire brigade or police”[CW] also indicated the need for a more integrated, user-centric notification system.

Participants from both groups recognised the importance of the functionalities in the current system and provided several suggestions for further improving the prototype.

\subsubsection{Content}
Both the interview and cognitive walkthrough participants found the app’s content useful, clear, and appropriate for disaster preparedness. The interview participants especially found the disaster guides' content meaningful, with one participant [INT] calling the Plan and Prepare section ``extremely appropriate", indicating that the app effectively supported readiness before disasters. Participants also praised the app’s use of concise, everyday language, noting that this made information accessible even to those with lower literacy levels. One elderly participant [INT] highlighted that functional illiteracy remains high in Australia, and that using familiar, simple words helped ensure inclusivity. Similarly, during cognitive walkthroughs, participants appreciated the clarity of the Take Action section—one low-digital-literacy user [CW] noted, “I like how the content is presented in Take Action as it is very clear for me with not too many choices.” These findings underscore that simplicity, directness, and readability are critical strengths of the application. This is particularly vital those who are not knowledgeable about disaster preparation or those who may be under stress during an emergency, where the ability to quickly comprehend and act on information can significantly impact safety and outcomes. 

The use of icons and illustrations also received generally positive feedback. Many participants described the visuals as clear, understandable, and not overly complicated. However, while some users found the icons left “no room for confusion,”[INT] others reported they “had to think twice,”[INT], suggesting minor inconsistencies in clarity that could be refined. The combination of short text and relevant visuals was appreciated for its ability to make complex evacuation or safety information more digestible [CW].

Despite these strengths, several participants raised concerns about clarity and depth of content. A number of users found some illustrations or layouts confusing, particularly within the evacuation section. Feedback such as “confusing illustration in evacuation section”[CW] and “issues on illustration of fire—what is the arrow?”[CW] pointed to the need for more descriptive visual aids and clearer graphical representations of actions. Similarly, some users preferred text over icons, expressing that “icons are superfluous; prefer to stick with text. The less there is, the faster it can be conveyed.”[CW] Another participant commented that “the icons used in Shelter in Place (e.g., wavy lines, fire extinguisher, shield) do not make sense”[CW]. These observations highlight the need to carefully balance visuals and text, especially for users with low digital literacy or age-related visual impairments.

Finally, participants identified areas where additional or more actionable content would strengthen the app. One participant [INT] observed that the evacuation section primarily listed dangers but not the corresponding responses, describing it as more appropriate for preparation rather than for real-time action during a catastrophic fire event. Another rural resident [INT] requested that the app include guidance for periods outside the fire season, such as early preparation and maintenance tasks. Together, these comments suggest expanding content coverage to support users across all phases of disaster management—before, during, and after an event.

Overall, participants found the content well-written and meaningful, with clear potential for real-world application. However, they also suggested improvements in visual design clarity and actionable content to enhance the system’s effectiveness in communicating critical disaster information to diverse and vulnerable communities.

\subsubsection{User Interface}
The feedback on the user interface (UI) from both the interviews and cognitive walkthroughs was mostly positive, with participants describing the app as simple, clean, and easy to use. Many appreciated the layout and structure, noting that navigation felt intuitive and that they could easily find their place in the app. Interview participants highlighted that the quick links at the end of pages as particularly useful for moving between sections, while the bottom navigation bar was found to be familiar and not problematic, even for elderly users. Interviewed Rural residents described it as convenient and practical. Several participants also appreciated the overall visual design, and one elderly participant mentioned that the chosen font, `Inter', was “clean, good looking,” and made the interface feel “simple and well laid out”[INT]. Participants from the cognitive walkthroughs also commented on the layout’s consistency, which they felt made the app easy to learn and navigate. The general view was that familiarity and predictability are essential in an emergency app, as they reduce hesitation and cognitive effort under stress and that this was successfully achieved in the prototype [CW]. 

However, participants also identified some usability challenges. In interacting with the map, some participants were unsure what specific elements meant. For example, the blue dot representing the current location was mistaken for another feature (“I presume the blue dot is Orange[the closest main town], is it?”)[INT], and the dotted lines for danger zones were interpreted as contour lines. This confusion was caused by the lack of a legend or scale[INT]. Some elderly participants described the app as “not immediately intuitive but easy to learn,” noting that they needed “a couple of goes to understand it” [INT]. In the cognitive walkthroughs, several users found that navigation sometimes required extra cognitive effort, especially when moving between the danger-rating pages and corresponding guides. Participants portraying rural or low-digital-literacy users[CW] were more likely to lose track of where they had come from or how to return to a previous section. One rural participant [CW] commented that the layout might be less suitable for some elderly users unless they are guided by someone familiar with the app.

Participants also shared several suggestions to make the interface more adaptive and easier to follow. One common idea was to automatically reorder the guide links based on the current fire danger rating—for example, showing ``Take Action" first when the risk is extreme [CW]. Others proposed linking the fire danger rating indicators directly to their relevant guide sections so users could act more quickly in emergencies[CW]. A few participants asked for clearer visual aids, such as a vertical scroll bar or a floating navigation tab, to help them track their reading progress through long pages[INT]. There were also aesthetic and layout preferences: some users wanted icons arranged horizontally beside text (rather than vertically) and suggested slightly smaller icon sizes to minimise scrolling[INT]. Several participants expressed a preference for text over icons altogether, saying that “the less there is, the faster it can be conveyed” [CW].

A few participants also highlighted the need for more responsive feedback. For instance, one colour-blind participant[CW] mentioned that the app should provide feedback when a new colour scale is selected, confirming that the choice was applied correctly. This type of responsive interaction was seen as critical for building trust and confidence, especially in safety-critical contexts.

Overall, participants found the app’s user interface visually clean, consistent, and learnable, but their feedback points to further improvement. They identified that enhancing navigational aids, clarifying map elements, improving icon consistency, and adding more feedback mechanisms would strengthen the app’s usability and inclusiveness, making it better suited to support users with diverse needs during high-stress emergency situations.

\subsubsection{Adaptivity} \label{sec:adaptivity}
Participants across both the interviews and CW recognised that the app’s adaptive features added real value to the user experience. Overall, they found the adaptive options easy to set up and use, and many appreciated the system's ability to personalise how they received and interacted with information. The initial profiling process was described as straightforward and not burdensome. One elderly participant[INT] commented that ``you can actually set it up so eventually you'll be able to become quite comfortable with it,” reflecting a sense of gradual confidence-building through familiarity. Similarly, participants appreciated that the app gave them a sense of control over how information was displayed, with one rural participant[INT] noting that the page mode option could help avoid losing focus when reading longer content: “scrolling to me, you can get quite lost when there’s a lot of information.” The feedback from a rural persona participant [CW] also confirmed that the system’s ability to function even with limited Internet connectivity was an important strength in emergency management contexts, as it enabled access to evacuation content offline. Participants with low digital literacy personas [CW] particularly appreciated the clarity of language and the concise instructions throughout the adaptive menus, describing the tone as easy to follow and unintimidating.

Despite this positive feedback, participants also showed some confusion and inconsistency in how the adaptive features were presented. Some users admitted they skipped certain options, such as disabling visual cues, because they “didn’t understand what it meant”[INT]. Others were unsure about the purpose of certain adaptive elements, such as page mode, because they felt they were not directly relevant to them, which made those features feel unnecessary or even distracting [INT]. Participants[CW] representing colour-blind users noted that the colour customisation options were not intuitive and that it was difficult to find combinations that worked reliably. While some participants felt the buttons and interactive elements were appropriately sized (“large enough to be easily tapped without error”), others described the text, particularly on the home page, as too small[CW]. This tension between readability and interface compactness highlighted the challenge of balancing accessibility and usability across different user groups.

Participants also provided several thoughtful suggestions for improving the adaptivity design. The strongest theme was the need for clearer explanations and feedback. Users wanted the app to explain what each adaptive option meant, either during setup or through short in-context help prompts. Several participants[CW] emphasised the need for confirmation when changes were applied, for example, “Your colour scale has been updated”, to make sure their choices had taken effect. Others[CW] asked for simpler terminology in menus, such as using ``Push Notifications" instead of simple “Push” as this was confusing for those with lower digital literacy. There were also requests for greater dynamic adaptivity, particularly for evacuation information. For example, one rural participant [CW] asked, “What if the fire cuts off the main road during my evacuation?” This pointed to the need for location-specific, real-time adaptivity to keep content relevant as situations evolve.

These show that while some adaptive features were well received, others could complicate the user experience. This highlights the need for clearer explanations and more intuitive design choices to ensure the adaptive features enhance rather than hinder user interaction. 

\subsubsection{Requirement Traceability Matrix (RTM)}

\begin{table}[ht]
\centering
\caption{Summary of validation outcomes by community group (R01-R67)}
\label{tab:validation-summary}
\scriptsize
\setlength{\tabcolsep}{4pt}
\renewcommand{\arraystretch}{1.3}
\begin{tabular}{lc ccc c ccc c c}
\toprule\noalign{\vspace{-2pt}}

\multirow{2}{*}{\textbf{Community group}}
  & \multirow{2}{*}{\textbf{Total}}
  & \multicolumn{4}{c}{\cellcolor{clrsummary}\textbf{Addressed in validation}}
  & \multicolumn{3}{c}{\cellcolor{clrsummary}\textbf{Positively validated}}
  & \multirow{2}{*}{\cellcolor{clrnoeval}}
  & \multirow{2}{*}{} \\

\noalign{\vspace{-3pt}}\cmidrule(lr){3-6}\cmidrule(lr){7-9}\noalign{\vspace{-2pt}}

  &
  & \cellcolor{clrsupported}\textbf{Sup.}
  & \cellcolor{clrpartial}\textbf{Part.}
  & \cellcolor{clrrevision}\textbf{Rev.}
  & \cellcolor{clrsummary}\textbf{\shortstack{Total/\\All reqs\\(\%)}}
  & \cellcolor{clrsupported}\textbf{Sup.}
  & \cellcolor{clrpartial}\textbf{Part.}
  & \cellcolor{clrsummary}\textbf{\shortstack{Total/\\All reqs\\(\%)}}
  & \cellcolor{clrnoeval} \textbf{\shortstack{Not\\Evaluated}}
  & \textbf{Reqs} \\

\noalign{\vspace{-3pt}}\midrule\noalign{\vspace{-2pt}}

\multicolumn{11}{l}{\cellcolor{clrgrphead}\textit{Individual community groups}} \\
\noalign{\vspace{-3pt}}\midrule\noalign{\vspace{-2pt}}

\faBlind\ Elderly
  & 13
  & \cellcolor{clrsupported}6
  & \cellcolor{clrpartial}5
  & \cellcolor{clrrevision}2
  & \cellcolor{clrsummary}13/13 (100\%)
  & \cellcolor{clrsupported}6
  & \cellcolor{clrpartial}5
  & \cellcolor{clrsummary}11/13 (85\%)
  & \cellcolor{clrnoeval}0
  & R01--R13 \\

\faKeyboard\ Low-Digital-Literacy
  & 16
  & \cellcolor{clrsupported}3
  & \cellcolor{clrpartial}9
  & \cellcolor{clrrevision}1
  & \cellcolor{clrsummary}13/16 (81\%)
  & \cellcolor{clrsupported}3
  & \cellcolor{clrpartial}9
  & \cellcolor{clrsummary}12/16 (75\%)
  & \cellcolor{clrnoeval}3
  & R14--R29 \\

\faTree\ Rural
  & 5
  & \cellcolor{clrsupported}2
  & \cellcolor{clrpartial}2
  & \cellcolor{clrrevision}0
  & \cellcolor{clrsummary}4/5 (80\%)
  & \cellcolor{clrsupported}2
  & \cellcolor{clrpartial}2
  & \cellcolor{clrsummary}4/5 (80\%)
  & \cellcolor{clrnoeval}1
  & R30--R34 \\

\faLowVision\ Colour-Blind
  & 10
  & \cellcolor{clrsupported}3
  & \cellcolor{clrpartial}2
  & \cellcolor{clrrevision}4
  & \cellcolor{clrsummary}9/10 (90\%)
  & \cellcolor{clrsupported}3
  & \cellcolor{clrpartial}2
  & \cellcolor{clrsummary}5/10 (50\%)
  & \cellcolor{clrnoeval}1
  & R35--R44 \\

\noalign{\vspace{-3pt}}\midrule\noalign{\vspace{-2pt}}

\multicolumn{11}{l}{\cellcolor{clrgrphead}\textit{Shared community groups}} \\
\noalign{\vspace{-3pt}}\midrule\noalign{\vspace{-2pt}}

\faBlind\ \faLowVision\ Elderly + Colour-Blind
  & 3
  & \cellcolor{clrsupported}0
  & \cellcolor{clrpartial}1
  & \cellcolor{clrrevision}0
  & \cellcolor{clrsummary}1/3 (33\%)
  & \cellcolor{clrsupported}0
  & \cellcolor{clrpartial}1
  & \cellcolor{clrsummary}1/3 (33\%)
  & \cellcolor{clrnoeval}2
  & R45--R47 \\

\faBlind\ \faTree\ Elderly + Rural
  & 9
  & \cellcolor{clrsupported}4
  & \cellcolor{clrpartial}4
  & \cellcolor{clrrevision}1
  & \cellcolor{clrsummary}9/9 (100\%)
  & \cellcolor{clrsupported}4
  & \cellcolor{clrpartial}4
  & \cellcolor{clrsummary}8/9 (89\%)
  & \cellcolor{clrnoeval}0
  & R48--R56 \\

\faTree\ \faKeyboard\ Rural + Low-Digital-Literacy
  & 5
  & \cellcolor{clrsupported}1
  & \cellcolor{clrpartial}0
  & \cellcolor{clrrevision}0
  & \cellcolor{clrsummary}1/5 (20\%)
  & \cellcolor{clrsupported}1
  & \cellcolor{clrpartial}0
  & \cellcolor{clrsummary}1/5 (20\%)
  & \cellcolor{clrnoeval}4
  & R57--R61 \\

\faBlind\ \faTree\ \faKeyboard\ Elderly + Rural + Low-Digital-Literacy
  & 6
  & \cellcolor{clrsupported}2
  & \cellcolor{clrpartial}3
  & \cellcolor{clrrevision}4
  & \cellcolor{clrsummary}6/6 (100\%)
  & \cellcolor{clrsupported}2
  & \cellcolor{clrpartial}3
  & \cellcolor{clrsummary}5/6 (83\%)
  & \cellcolor{clrnoeval}0
  & R62--R67 \\

\noalign{\vspace{-3pt}}\midrule\noalign{\vspace{-2pt}}

\textbf{Total}
  & \textbf{67}
  & \cellcolor{clrsupported}\textbf{21}
  & \cellcolor{clrpartial}\textbf{26}
  & \cellcolor{clrrevision}\textbf{12}
  & \cellcolor{clrsummary}\textbf{56/67 (84\%)}
  & \cellcolor{clrsupported}\textbf{21}
  & \cellcolor{clrpartial}\textbf{26}
  & \cellcolor{clrsummary}\textbf{47/67 (70\%)}
  & \cellcolor{clrnoeval}\textbf{11}
  & \\

\noalign{\vspace{-3pt}}\bottomrule
\end{tabular}
\end{table}

Table~\ref{tab:validation-summary} summarises how the 67 requirements were addressed during validation. The detailed requirement traceability matrix is available in Appendix E. Overall, 56 requirements (84\%) were addressed through the prototype-based evaluation, and 47 requirements (70\%) received positive validation evidence. The remaining 11 requirements were not evaluated because they related to aspects that could not be meaningfully tested in a Figma prototype, including performance, connectivity, audio output, alt-text, and crash recovery. We therefore treat these as prototype limitations rather than design failures. These requirements will need to be evaluated in future work using a functionally deployed system.

The validation coverage varied across the community groups. The elderly group and elderly + rural groups' requirements achieved the highest coverage, with all requirements addressed during validation. This is expected, as these groups were directly represented in the interview study. The colour-blind group also had high coverage, with 90\% of requirements addressed, but had the lowest positive validation rate (50\%). This was mainly due to the usability issues participants identified with the colour customisation features, as discussed in Section~\ref{sec:adaptivity}.

Across all groups, 12 requirements were identified as needing revision. The highest concentrations were in the colour-blind group and the shared elderly + rural + low-digital-literacy group, with four requirements in each. These findings should not be interpreted as design failures. Rather, they indicate that the validation process produced actionable feedback that can guide the next iteration of the prototype. Among the shared requirement groups, rural + low-digital-literacy had the lowest addressed rate (20\%). This was largely because several requirements in this group focused on performance and connectivity, which could not be meaningfully evaluated using a Figma prototype.

This traceability matrix shows that the prototype operationalised most of the identified requirements, and stakeholder feedback provided useful evidence of their relevance and limitations. The unevaluated requirements also highlight where prototype-based validation is insufficient, especially for requirements that depend on implementation-level behaviour or deployment conditions. We discuss these implications further in the Discussion section.

\section{Discussion}
\textbf{Reflections on Requirements for Vulnerable Communities: } 
The four vulnerable communities involved in this study (elderly, individuals with low digital literacy, rural residents, and colour-impaired) showed that, while each group had distinct needs, many of the underlying design principles that supported their inclusivity overlapped. For example, simplicity and reduced cognitive load were important for both the elderly and low-digital-literacy communities. However, subtle differences existed in how these requirements should be operationalised—the elderly valued gradual introduction of features and larger visual elements, whereas users with low digital literacy benefitted more from contextual prompts and consistent feedback. Similarly, while colour-impaired communities valued high contrast ratios and customisable colour scales, interestingly, several of these requirements also benefited elderly users, whose visual acuity may be reduced with age. These examples show that designing for vulnerability can indeed enhance usability for all. 

\textbf{Adaptivity in critical systems: } Across all groups, participants valued the ability to adapt and personalise the system to suit their individual needs. They particularly appreciated having control over visual and interactional aspects such as colour, font, and navigation modes. However, the participatory evaluation also revealed that adaptive features must be clearly explained and easy to modify. When users were unsure of what a feature did, they either avoided using it or applied it incorrectly—both of which could lead to hesitation or inaction during critical moments. These findings show that while adaptivity helps meet the varying requirements of user groups, for critical systems they should also give considerable attention to how understandable and controllable those mechanisms are for end users.

\textbf{Prototyping for requirements modelling and validation: } This study used prototyping to model and validate our identified requirements. Unlike conventional modelling approaches that rely on formal notations and abstract representations, prototyping helped us to prioritise visual/experiential engagement. We found that when interacting with vulnerable communities, this was most effective because it allowed us to translate abstract requirements into tangible, interactive representations. The visual and experiential nature of the prototype helped participants better understand the intended functionality and design, which, in turn, allowed them to provide more effective feedback. For example, users with low digital literacy and older adults were able to comment on layout, readability, and navigation aspects more easily when they could interact with the prototype. Similarly, participants with colour-vision impairments could directly evaluate the effectiveness of colour scales and contrast ratios once visualised, which would not have been possible through text alone. Therefore, prototyping can be recommended as one of the best methods for participatory requirements modelling and validation.

\textbf{Inclusivity of critical systems: }The process followed in this study illustrates how inclusivity can be embedded from the earliest stages of RE, rather than treated as a late-stage usability concern. While a single case study cannot establish this as a universal principle, the findings suggest that deferring inclusivity to later stages of design risks producing systems that fail the communities that depend on them most, particularly in safety-critical contexts where the consequences of such failures extend beyond inconvenience. System developers need to understand that in these systems, inclusivity is not just about accessibility.  Instead, it is also about making the systems reliable and resilient so that ``no one is left behind" during safety-critical situations. Inclusivity must therefore be introduced from the very beginning of the RE process and continually refined through community-informed feedback. The process demonstrated in this study, from eliciting design guidelines in the literature to specifying requirements, operationalising them through a prototype, and validating them with stakeholders, offers a repeatable pathway for socially responsible RE in safety-critical domains.

\subsection{Recommendations and A Call to Action}
We propose some recommendations presented in Table \ref{tab_recommendations} for system designers, requirements engineers, and policymakers. These recommendations aim to ensure a more human-centric, participatory RE process that considers vulnerable communities. They also help ensure that critical systems are accessible for all, especially during safety-critical situations.

\noindent
\captionof{table}{Recommendations for stakeholder groups to ensure more Human-centric RE}
\label{tab_recommendations}
\begin{tabularx}{\columnwidth}{ l  X }
    \toprule
    \textbf{Stakeholder Group} & \textbf{Recommendations} \\
    \midrule
   \multirow{2}*{Requirements} & 1. Treat human-centricity as a safety-critical property that can affect resilience and user trust \\
      & 2. Maintain traceability between inclusive guidelines, requirements, and validation evidence\\
    Engineers  & 3. Combine qualitative participatory insights with formal requirements documentation (e.g., Linking Interviews and Cognitive Walkthrough findings to requirements via RTMs)\\
   \hline & \\[-2ex]
   \multirow{4}*{System Designers} & 1. Integrate human-centricity checks at the requirements elicitation stage, not as part of later usability testing.\\
       & 2. Prioritise adaptable and explainable interfaces over one-size-fits-all accessibility solutions.\\
    \&  Developers  & 3. Employ participatory validation cycles (e.g., short, persona-driven walkthroughs) when access to real users is limited.\\
    \hline & \\[-2ex]
    \multirow{2}*{Policymakers \&} & 1. Encourage frameworks and funding mechanisms that require human-centric RE in safety-critical domains.\\
     Funding Bodies & 2. Recognise human-centricity as resilience; policies should treat human-centricity not as an ethical add-on but as an operational dimension of resilience in critical systems.\\
   \bottomrule
\end{tabularx}

\begin{figure}[h]
\centering
\includegraphics[width=0.85\textwidth, trim={6cm 2cm 5cm 0.5cm},clip]{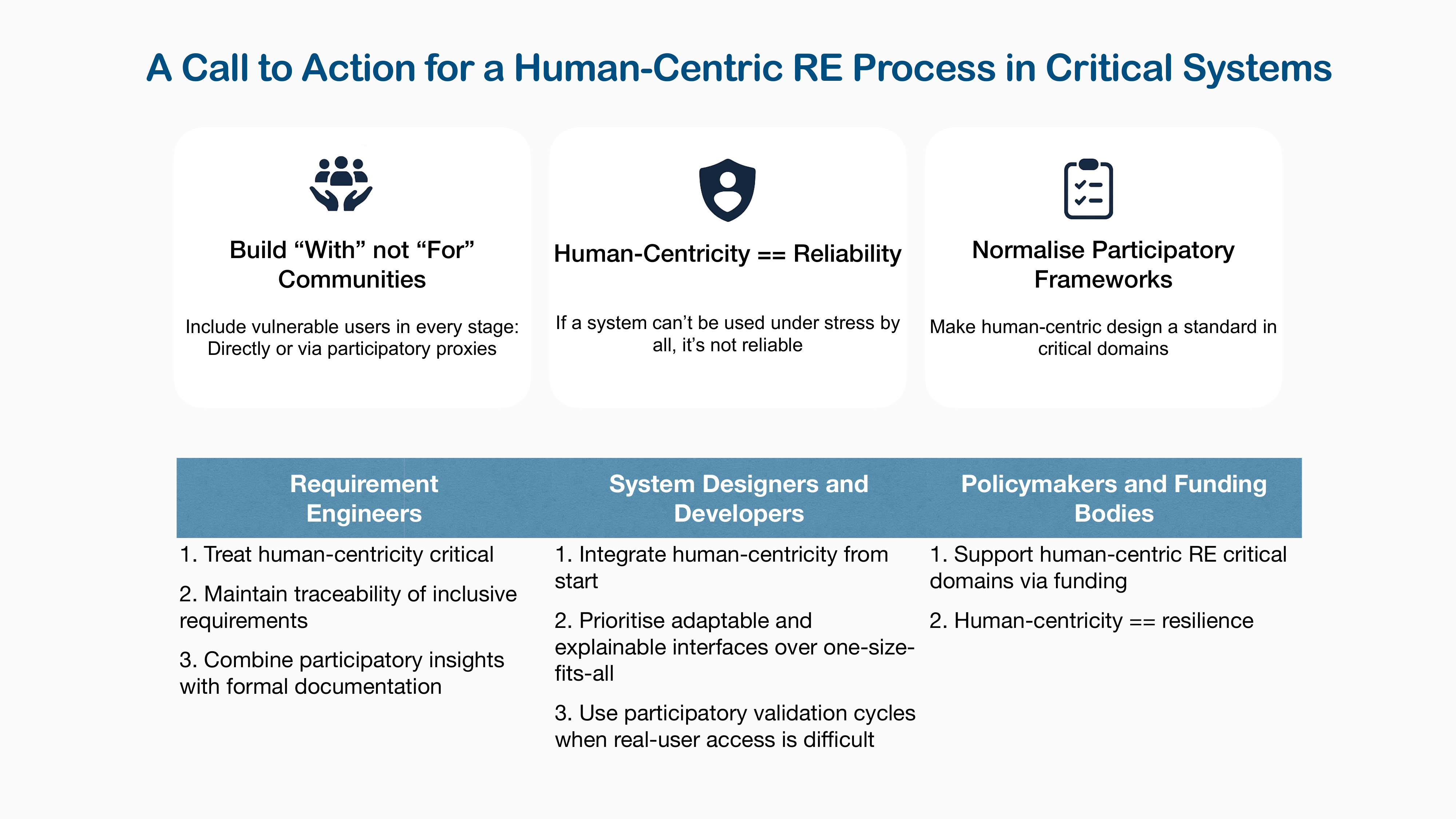}
\caption{An Action Plan}
\label{fig_actionPlan}
\end{figure}

We bring forward a call to action for critical systems. Critical systems exist to protect people, yet research has shown that the most vulnerable communities are consistently underrepresented in the design processes that shape these systems \cite{martins2020, tizard2024}, leaving them at greater risk when these systems fail. As disasters intensify and technology becomes central to emergency response, human-centricity is no longer optional. It is a fundamental component of safety. This study has shown that participatory requirements engineering can bridge that gap, translating community needs into implementable design requirements that make systems truly life-saving. The way forward is clear, and we have presented a summarised version of our recommendations and a plan for future critical system design in Figure \ref{fig_actionPlan}.

\section{Threats to Validity}
\subsection{Construct Validity}
The guidelines analysed in this study were derived from prior participatory work, helping to ensure that the constructs were grounded in authentic user needs. To support the selection process, three authors independently reviewed the relevant literature and then cross-checked the extracted guidelines to identify overlaps and resolve inconsistencies. When multiple guidelines addressed similar design concerns, we prioritised those supported by stronger methodological detail, larger participant samples, or richer descriptions of the guideline and its rationale. This reduced the likelihood of arbitrary inclusion and helped ensure that the selected guidelines reflected the strongest available evidence for each design concern. However, as the review was targeted rather than systematic, it is possible that some relevant guidelines were not captured. In addition, the selection criteria relied on researcher judgement rather than a formal review protocol. Some bias in guideline inclusion may therefore remain. Future work could strengthen this evidence base through a systematic review of inclusive design guidelines for vulnerable communities.

\subsection{Internal Validity}
While the adaptive prototype was directly informed by the elicited and specified requirements, participants' feedback may have been influenced by their prior familiarity with similar applications or by the demonstration format used during the sessions. To minimise this, both the interviews and cognitive walkthroughs followed structured prompts and task-based evaluation procedures. 

The interview study involved six participants, including two older adults and four rural residents, and therefore provided direct feedback for these two groups. For the low-digital-literacy and colour-blind groups, we used persona-based cognitive walkthroughs as a complementary validation method. Personas and cognitive walkthroughs are well-established methods in HCI and software engineering for evaluating interaction scenarios, particularly when direct access to specific user groups is constrained \cite{burnett2016gendermag,spencer2000streamlined}. This approach allowed us to systematically examine how requirements may support users with different capabilities and constraints. Nevertheless, persona-based evaluation does not fully replace direct participation from the represented communities. Future work should therefore extend the validation with direct participation from low-digital-literacy and colour-blind community members, along with larger samples across all four groups. 

A more rigorous evaluation could also involve a comparative study with two prototype versions, one designed with and one without the inclusive requirements, to provide stronger empirical evidence of the requirements' impact on usability and accessibility outcomes for vulnerable communities. 

Additionally, the four personas used in the cognitive walkthroughs represented one persona per vulnerable group, which does not effectively capture the variation within each group. For example, older adults vary considerably in digital literacy, physical ability, and prior technology experience, and a single elderly persona cannot represent this diversity. Future work should develop richer persona sets that reflect intra-group variation to strengthen the representativeness of cognitive walkthrough evaluations.

\subsection{External Validity}
This study focused on four vulnerable user groups: older adults, low-digital-literacy users, rural users, and colour-blind users. Other forms of vulnerability, such as linguistic diversity, cognitive impairments, and socioeconomic constraints, were outside the scope of this study and should be examined in future work. 

The prototype and validation were also situated in an Australian context, with bushfire early warning selected as the case domain because of its local relevance. This geographic and cultural specificity may influence how the findings transfer to other disaster contexts or countries. However, the inclusive design guidelines underpinning this work were drawn from broader usability and accessibility literature, rather than from disaster-specific guidance alone. This was necessary because inclusive design guidance for disaster early warning systems is not yet established as a mature body of work. Translating these broader guidelines into disaster-relevant requirements was therefore a significant contribution of this study. Importantly, as evidenced by the 67 requirements presented in Appendix C, none of the derived requirements are specific to bushfire or any other particular hazard type; they are expressed in terms applicable to any disaster early warning or safety-critical system. The translation process demonstrated here could therefore be repeated for other critical domains without modification to the requirement set itself. While the validation findings are specific to the bushfire prototype evaluated in this study, the overall process is likely to be transferable beyond this context. Future studies should replicate this process in other disaster contexts and in other user-facing critical domains, such as healthcare, transportation, and defence, to examine its broader transferability.

\section{Conclusion}
This study illustrated how a human-centred RE process can be applied within a critical domain to make systems more inclusive and socially responsible. Using disaster early warning systems as the case context, we showed how participatory and human-centric RE practices can uncover requirements that traditional approaches often overlook. The findings highlight that human-centric requirements not only improve access for vulnerable groups but also enhance usability and resilience for all users, emphasising the argument that human-centricity should be a core quality attribute for critical systems.

Our findings demonstrate that when the requirements of vulnerable communities are considered early on, the resulting systems become more understandable and accessible to everyone. The participatory validation activities also revealed that users value adaptive interfaces and information formats that build on their diverse capabilities and contexts. However, the results also revealed that this adaptivity must always remain explainable and controllable, especially in high-stakes environments where misunderstanding or hesitation can have severe consequences. 

Based on the findings of this case study, we argue that human-centricity should no longer be treated as an ethical or peripheral concern in critical systems but as an operational dimension of safety and reliability. Future research can extend the human-centric RE process to other safety-critical domains such as healthcare, defence, and transportation. These should also explore additional participatory methods to involve vulnerable user groups throughout the RE lifecycle. Ultimately, human-centred RE offers a pathway towards systems that not only function correctly but also serve all members of society with equity.

\section*{Acknowledgments}
Grundy is supported by ARC Laureate Fellowship FL190100035

\appendix

\section*{Appendix A: Guidelines Specific to a Single Vulnerable Community} 
Link to the detailed guidelines can be found here: \url{https://github.com/anukmd/human_centric_RE_repo/blob/main/Unique_Guidelines.pdf}

\section*{Appendix B: Guidelines Shared between Vulnerable Communities} 
Link to the detailed guidelines can be found here: \url{https://github.com/anukmd/human_centric_RE_repo/blob/main/Shared_Guidelines.pdf}

\section*{Appendix C: Requirements for Vulnerable Community} 
Link to the detailed guidelines can be found here: \url{https://github.com/anukmd/human_centric_RE_repo/blob/main/Identified_Requirements.pdf}

\section*{Appendix D: Personas used in Cognitive Walkthrough} \label{appendix_personas}
Link to the detailed guidelines can be found here: \url{https://github.com/anukmd/human_centric_RE_repo/blob/main/Personas.pdf}

\section*{Appendix E: Requirement Traceability Matrix} \label{appendix_rtm}
Link to the detailed matrix can be found here: \url{https://github.com/anukmd/human_centric_RE_repo/blob/main/Requirement%20Traceability%20Matrix.pdf}

\nocite{ColorBli89_online,Colorbli3_online,Designin57_online,HowToDes35_online,Howcanmo84_online,UserExpe81_online,WebConte89_online,czaja1990computer,gilbert2019inclusive,gomez2023design,groves2000web,hardy2019designing,maphosa2021covid,norman2013design,roberts2017review,ruzic2017universal,shneiderman2021designing,srivastava2021actionable,tak2022,tan_modified_2020,Whataret15_online,zaina2022preventing}

\bibliographystyle{elsarticle-num-names} 
\bibliography{references_jss}

\end{document}